\documentclass[aps,twocolumn,prb,floatfix,superscriptaddress,showpacs,lengthcheck,hyperref,citeautoscript]{revtex4-1}
\usepackage{amsmath}
\usepackage{amssymb}
\usepackage{graphicx}
\usepackage{bm}
\usepackage{color}
\usepackage{times}
\usepackage{MnSymbol}
\usepackage{verbatim}

\newcommand{\smeq}{\! = \!}

\newcommand{\smmi}{\! - \!}

\newcommand{\ve}{\varepsilon}

\newcommand{\bea}{\begin{eqnarray}}
\newcommand{\eea}{\end{eqnarray}}
\newcommand{\be}{\begin{equation}}
\newcommand{\ee}{\end{equation}}

\newcommand{\la}{\langle}
\newcommand{\ra}{\rangle}
\newcommand{\ket}[1]{| #1 \rangle}

\begin{document}

\title{Floquet bound states around defects and adatoms in graphene}
\author{D. A. Lovey}
\affiliation{Centro At{\'{o}}mico Bariloche and Instituto Balseiro,
Comisi\'on Nacional de Energ\'{\i}a At\'omica, 8400 Bariloche, Argentina}
\affiliation{Consejo Nacional de Investigaciones Cient\'{\i}ficas y T\'ecnicas (CONICET), Argentina}
\author{Gonzalo Usaj}
\affiliation{Centro At{\'{o}}mico Bariloche and Instituto Balseiro,
Comisi\'on Nacional de Energ\'{\i}a At\'omica, 8400 Bariloche, Argentina}
\affiliation{Consejo Nacional de Investigaciones Cient\'{\i}ficas y T\'ecnicas (CONICET), Argentina}
\author{L. E. F. Foa Torres }
\affiliation{Departamento de F\'{\i}sica, Facultad de Ciencias F\'{\i}sicas y Matem\'aticas, Universidad de Chile, Santiago, Chile}
\author{C. A. Balseiro}
\affiliation{Centro At{\'{o}}mico Bariloche and Instituto Balseiro,
Comisi\'on Nacional de Energ\'{\i}a At\'omica, 8400 Bariloche, Argentina}
\affiliation{Consejo Nacional de Investigaciones Cient\'{\i}ficas y T\'ecnicas (CONICET), Argentina}

\begin{abstract}
Recent studies have focused on laser-induced gaps in graphene which have been shown to have a topological origin, thereby hosting robust states at the sample edges.
While the focus has remained mainly on these topological chiral edge states, the Floquet bound states around defects lack a detailed study.
In this paper we present such a study covering large defects of different shape and also vacancy-like defects and adatoms at the dynamical gap at $\hbar\Omega/2$ ($\hbar\Omega$ being the photon energy). Our results, based on analytical calculations as well as numerics for full tight-binding models, show that the bound states are chiral and appear in a number which grows with the defect size. Furthermore, while the bound states exist regardless the type of the defect's edge termination (zigzag, armchair, mixed), the spectrum is strongly dependent on it. In the case of top adatoms,  the bound states quasi-energies depend on the adatoms energy. The appearance of such bound states might open the door to the presence of  topological effects on the bulk transport properties of dirty graphene.
\end{abstract}
\date{\today}
\pacs{73.22.Pr; 73.20.At; 72.80.Vp; 78.67.-n}
\maketitle

\section{Introduction}
Driving a material out of equilibrium offers interesting paths to alter and tune its electrical response. A prominent example is the generation of light-induced topological properties~\cite{Oka2009,Lindner2011,Kitagawa2010}, e.g. illuminating a material like graphene to transform it in a Floquet topological insulator (FTI). Very much as ordinary topological insulators (TI),~\cite{Hasan2010,Kane2005,Ando2013,Ortmann2015} FTIs have a gap in their bulk (quasi-) energy spectrum---being then a bulk insulator---and their Floquet-Bloch bands are characterized by non-trivial topological invariants.~\cite{Rudner2013,Kitagawa2010,Gomez-Leon2013} In addition, and despite some important differences with TIs,~\cite{Rudner2013,Carpentier2015} FTIs show a bulk-boundary correspondence and hence host chiral/helical states at the sample boundaries.

The emergence of such non-equilibrium properties has been intensively investigated in recent years in a variety of systems including graphene~\cite{Calvo2011,Zhou2011,Kitagawa2011,Iurov2012,SuarezMorell2012,Perez-Piskunow2014,Usaj2014a,Beugeling2014,Perez-Piskunow2015} and other $2$D materials~\cite{Sie2014,Lopez}, normal insulators,~\cite{Lindner2011,Farrell2015} coupled Rashba wires,~\cite{Klinovaja2015} photonic crystals,~\cite{Rechtsman2013} cold atoms in optical lattices,~\cite{Goldman2014,Choudhury2014,Bilitewski2014,Dasgupta2015,DAlessio2014,Goldman2015,Mori2014} topological insulators,~\cite{Dora2012,Wang2013a,Calvo2015,DalLago2015,Gonzalez2016} and also classical systems~\cite{Fleury2015}.
The research interest has focused in many different aspects of the problem such as the characterization of the  edge states~\cite{Perez-Piskunow2014,Usaj2014a}, different signatures in magnetization and tunneling~\cite{Fregoso2014,Dahlhaus2014}, the proper invariants entering the bulk-boundary correspondence~\cite{Rudner2013,Ho2014,Carpentier2015,Perez-Piskunow2015}, their statistical properties~\cite{Dehghani2014,Liu2014}, the role of interactions and dissipation~\cite{Seetharam2015,Iadecola2015,Dehghani2014,Dehghani2015,Genske2015} and the associated two-terminal~\cite{Gu2011,Kundu2014} and multiterminal (Hall) conductance both in the scattering~\cite{FoaTorres2014} and decoherent regimes~\cite{Dehghani2015}.
So far, however, the experimental confirmation of the presence of such edge states has only been achieved in photonic crystals.~\cite{Rechtsman2013} Nonetheless, in condensed matter systems the Floquet induced gaps have already been observed at the surface of a topological insulator (Bi$_2$Se$_3$) by using time and angle resolved photoemission spectroscopy (tr-ARPES)~\cite{Wang2013a}. More recently, effective Floquet Hamiltonians were realized in cold matter systems.\cite{Jotzu2014}

\begin{figure}[b]
\includegraphics[width=0.9\columnwidth]{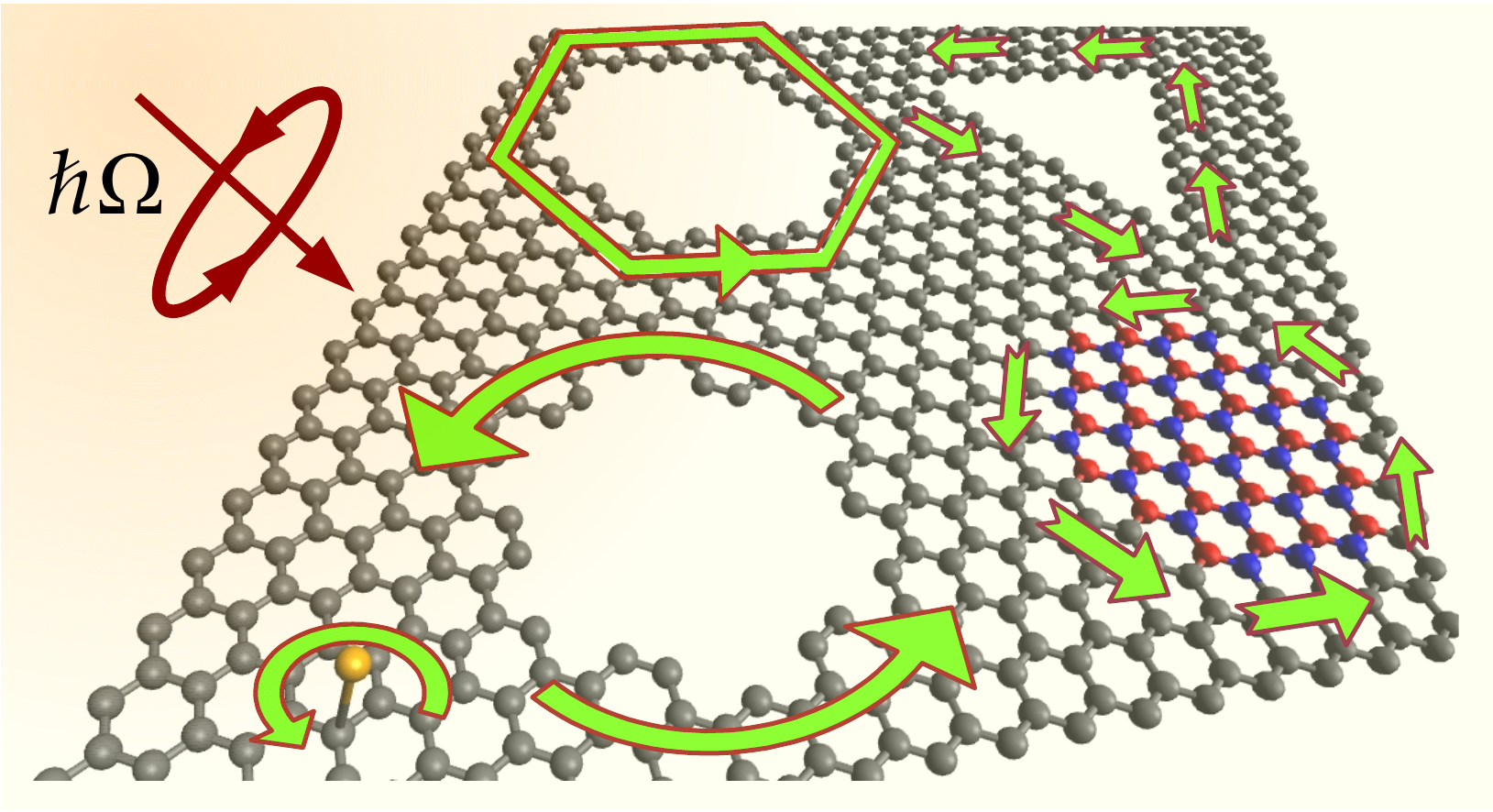}
\caption{(color online) Scheme of irradiated graphene with  different defects on the graphene lattice: holes, adatoms or regions with a staggering sublattice potential. The arrows indicate the chirality of the probability currents associated to the Floquet bound states around the defects.}
\label{scheme}
\end{figure}

Despite the intense research on FTIs, most of the studies address pristine samples. Besides occurring naturally in any sample, defects will also host Floquet bound states when the sample is illuminated. If the defects are extended, the presence of the associated Floquet bound states might allow for new experiments probing them. This motivates our present study. Specifically, taking laser-illuminated graphene as a paradigmatic example of a FTI, we study Floquet bound states around defects in the bulk of a sample. We show that chiral states circulate around holes or multi-vacancy defects of different shapes and lattice terminations (zigzag, armchair or mixed) like the ones showed in Fig. \ref{scheme}. The properties of these states (quasi-energies and their scaling with the system parameters, associated probability currents, etc.) are characterized using both numerical simulations, by means of a tight-binding model, and analytical approaches, by solving the appropriate low energy Dirac Hamiltonian in a reduced Floquet space. Quite interestingly, these bound states persist even in the limit of a single vacancy defect. Furthermore, bound states are found around adatoms that sit on top of a C atom (like H or F, for instance). 

While the presence of Floquet bound states around vacancy-like defects or adatoms might jeopardize the experimental observation of laser-induced gaps, they could, on the other hand, also open the route towards the observation of  interesting topological transport phenomena in dirty bulk samples by changing localization or percolation properties, for instance. 

The rest of the paper is organized as follows. First, we introduce our low energy model and the associated analytical Floquet solutions (Sec. \ref{model}). Several particular cases are presented in section \ref{BoundaryConditions}, namely, large holes with  zigzag or armchair edge terminations, as well as defects consisting of regions with a staggered potential. The chiral nature of the currents associated to the bound states is discussed in Sec. \ref{chiral_current}. In Sec. \ref{TBM} we compare our solutions with numerical calculations on a tight-binding model. The case of point like defects such as vacancies or adatoms is presented in Sec. \ref{adatom_sec}. We finally conclude in Sec. \ref{conclu}.

\section{The low energy model and the Floquet solution}
\label{model}
Let us consider an irradiated graphene sample with a single defect. Since the bound states we want to describe are  topological in origin,~\cite{Perez-Piskunow2014,Usaj2014a,Perez-Piskunow2015} the specific form or nature of the defect (see Fig. \ref{scheme}) is irrelevant for probing their existence---though the details of the quasi-energy spectrum and the particular form of the wave-functions will depend on it. To simplify the discussion we will start by assuming that the defect potential does not mix the different graphene valleys (Dirac cones)---this assumption will be relaxed when discussing particular examples. Hence, the low energy behavior around both cones can be described by a Hamiltonian given by
\begin{equation}
\hat{\cal{H}}(t)=v_F\,\bm{\sigma}\cdot\left(\bm{p}+
\frac{e}{c}\bm{A}(t)\right)+V(\bm{r})\,, 
\label{laser-graphene-Hamiltonian} 
\end{equation}
if we use the isotropic representation where the $K$ and $K'$ cones are described by the wave-functions $\psi_K(\bm{r},t)=\{\psi_A(\bm{r},t),\psi_B(\bm{r},t)\}^\mathrm{T}$ and $\psi_{K'}(\bm{r},t)=\{-\psi'_B(\bm{r},t),\psi'_A(\bm{r},t)\}^\mathrm{T}$, respectively.
Here $v_{F}\simeq 10^{6} $~m/s denotes the Fermi velocity, $\bm{\sigma}=(\sigma_x,\sigma_y)$ represents the Pauli matrices describing the pseudo-spin degree of freedom (sites $A$ and $B$ of the honeycomb lattice), $e$ is the absolute value of the electron charge, $c$ is the speed of light and $\bm{A}(t)=\mathrm{Re}\left\{\bm{A}_0e^{\mathrm{i}\Omega t}\right\}$ the vector potential of the electromagnetic field (a plane wave incident perpendicularly to the graphene sheet). 
The associated electric field is then $\bm{E}(t)=-(1/c)\partial_t \bm{A}(t)$ so that $|\bm{E}|=E_0=(\Omega/c) |\bm{A}_0|$.
It is important to emphasize that while we will refer to graphene from hereon, our results apply to any massless Dirac fermion system described by Eq. (\ref{laser-graphene-Hamiltonian}).

Since for solving the time-dependent Schr\"odinger equation we will take advantage of the Floquet formalism~\cite{Shirley1965,Sambe1973} used to deal with time dependent periodic Hamiltonians, it is instructive to briefly introduce its basic ideas (for a more extensive general reviews we refer to Refs. [\onlinecite{Grifoni1998a}] and [\onlinecite{Kohler2005}]).
Floquet theorem guarantees the existence of a set of solutions of the form $\ket{\psi_{\alpha}(t)}=\exp(-i\varepsilon_{\alpha}t/\hbar)\ket{\phi_{\alpha}(t)}$ where $\ket{\phi_{\alpha}(t)}$ has the same time-periodicity as the Hamiltonian, $\ket{\phi_{\alpha}(t+T)}=\ket{\phi_\alpha(t)}$ with $T=2\pi/\Omega$.\cite{Shirley1965,Grifoni1998a} The Floquet states $\ket{\phi_{\alpha}}$ are the solutions of the equation
\begin{equation}
\hat{\cal{H}}_F\ket{\phi_{\alpha}(t)} =\varepsilon_{\alpha}\ket{\phi_{\alpha}(t)}\,,
\label{floquetEq}
\end{equation}
where $\hat{\cal{H}}_F=\hat{\cal{H}}-i\hbar \partial_t$ is the Floquet Hamiltonian and $\varepsilon_{\alpha}$ the quasi-energy. 
Using the fact that the Floquet eigenfunctions are periodic in time, it is customary to introduce an extended $\cal{R}\otimes \cal{T}$ space (the Floquet or Sambe space\cite{Sambe1973}), where $\cal{R}$ is the usual Hilbert space and $\cal{T}$ is the space of periodic functions with period $T$.
A convenient basis of $\cal{R}\otimes \cal{T}$ can be built from the product of an arbitrary basis of $\cal{R}$ (the eigenfunctions $\ket{a_n}$ of the time-independent part of the Hamiltonian, for instance) and the set of orthonormal functions $e^{i m\Omega t}$, with $m=0,\pm1,\pm2, ...$ that span $\cal{T}$. Then, 
\begin{equation}
\label{fourier}
\ket{\phi_{\alpha}(t)}=\sum_{m=-\infty}^\infty \ket{u_m^\alpha}\,e^{i m\Omega t}\,,
\end{equation}
or, in a vector notation in  $\cal{R}\otimes \cal{T}$,
\begin{equation}
\ket{\phi_{\alpha}}=\{\cdots,\ket{u_1^\alpha},\ket{u_0^\alpha},\ket{u_{-1}^\alpha},\cdots\}^\mathrm{T}\,.
\label{vector}
\end{equation}
Here, $\ket{u^\alpha_m}=\sum_n  \mathcal{B}_{mn}^\alpha\ket{a_n}$ are linear combinations of the basis states of $\cal{R}$.
Written in this basis, $\hat{\cal{H}}_F $ is a time-independent infinite matrix operator with Floquet replicas shifted by a diagonal term $m\hbar\Omega$  and coupled by the radiation field with the condition, for pure harmonic potentials, that $\Delta m=\pm1$.

In the absence of any defect, the Floquet spectrum presents dynamical gaps at different quasi-energies\cite{Oka2009,Usaj2014a,Perez-Piskunow2015}. Here, we will focus on the gap, of order $\eta \hbar\Omega$, that appears at $\varepsilon\sim\hbar\Omega/2$ and look for bound states inside it. 
Since we will only consider the limit $\eta=v_F eA_0/c\hbar\Omega\ll1$, it is sufficient to restrict the Floquet Hamiltonian to the $m=0$ and $m=1$ subspaces (or replicas) for the analytical calculations---the numerical results can retain a larger number ($N_\mathrm{FR}$) of replicas if necessary. As discussed in Refs. [\onlinecite{Usaj2014a}] and [\onlinecite{Perez-Piskunow2015}], this restriction is enough to get the main features of the energy dispersion and the Floquet states when $\eta \ll 1$.

The reduced Floquet Hamiltonian describing states near $\varepsilon\sim\hbar\Omega/2$ then corresponds to
\begin{equation}
\tilde{\mathcal{H}}_F=\left(
\begin{array}{cccc}
\hbar\Omega & v_F p_- & 0 & 0\\
v_F p_+&\hbar\Omega &  \frac{v_Fe}{c}A_0 & 0 \\
0& \frac{v_Fe}{c}A_0 &0& v_F p_- \\
0&0 & v_F p_+ & 0\end{array}
\right)\,,
\label{matrix_k}
\end{equation}
with $p_{\pm}=p_x\pm i p_y=-i \hbar(\partial_x\pm i \partial_y)$. The Floquet wave-function has the form 
\be
\label{wavefunction}
\phi(\bm{r})=\{[u_{1A}(\bm{r}),u_{1B}(\bm{r})],[u_{0A}(\bm{r}),u_{0B}(\bm{r})]\}^\mathrm{T}\,.
\ee
It is straightforward to see that $\tilde{\mathcal{H}}_F\phi(\bm{r})=\varepsilon \phi(\bm{r})$ implies that
\bea
\nonumber
u_{1A}(\bm{r})&=&-\frac{v_F}{\hbar\Omega-\ve}\,p_-u_{1B}(\bm{r}),\\
u_{0B}(\bm{r})&=&\frac{v_F}{\ve}\,p_+u_{0A}(\bm{r})\,,
\label{eq1}
\eea
and hence only two functions, $u_{0A}(\bm{r})$ and $u_{1B}(\bm{r})$, have to be found. These functions satisfy
\bea
\nonumber
\left(-\frac{v_F^2}{\hbar\Omega-\ve}\,p^2+\hbar\Omega-\ve\right)u_{1B}(\bm{r})&=&-\frac{v_Fe}{c}A_0u_{0A}(\bm{r}),\\
\left(\frac{v_F^2}{\ve}\,p^2-\ve\right)u_{0A}(\bm{r})&=&-\frac{v_Fe}{c}A_0u_{1B}(\bm{r})\,,
\label{eq_2x2}
\eea
where $p^2=p_+p_-=p_-p_+$. 

Because we are interested in describing the effect of a defect---which breaks the translational invariance of the systems---, it is useful to change at this point to a polar coordinate system,  $r$ and $\varphi$, centered at it. 
In terms of these variables we have,
\bea
\nonumber
p_\pm&=&-i \mathrm{e}^{\pm i\varphi}\hbar\left(\partial_r\pm i\frac{1}{r}\partial_\varphi\right),\\
p^2&=&-\hbar^2\left(\partial^2_r+\frac{1}{r}\partial_r+\frac{1}{r^2}\partial^2_\varphi\right)\,.
\eea
Similarly, as in the case of local defects in ordinary TI~\cite{Lu2011,Shan2011}, the solutions of Eq. (\ref{eq_2x2}) can be written as  $u_{1B}(\bm{r})=\mathrm{e}^{ i l\varphi}f(k_0r)$ and $u_{0A}(\bm{r})=\mathrm{e}^{ i l\varphi}g(k_0r)$ with $l$ an integer number. 
This follows from the fact that $[\tilde{\mathcal{H}}_F,\mathcal{L}]=0$, where 
\be
	\mathcal{L} = \left(-i\hbar\partial_\varphi \otimes \sigma_0 + \frac{\hbar}{2}\sigma_z \right)\otimes \tau_0 + \frac{\hbar}{2} \sigma_0\otimes\tau_z 
\ee
and $\mathcal{L}\, \phi(\bm{r}) = \hbar l\, \phi(\bm{r})$, where $\phi(\bm{r})$ is given by Eq. (\ref{wavefunction}).
In order to proceed further  we define the adimensional parameters
\be
\mu=\frac{\ve}{\hbar\Omega/2}-1\,,\qquad k_0=\frac{\Omega}{2v_F}\,,\qquad \xi=k_0 r\,.
\ee
With this notation, the equations for $f(\xi)$ and $g(\xi)$ become 
\bea
\label{eqreduced}
\nonumber
\left[\left(\partial^2_\xi+\frac{1}{\xi}\partial_\xi-\frac{l^2}{\xi^2}\right)+(1-\mu)^2\right]f(\xi)&=&-2\eta(1-\mu)\, g(\xi),\\
\nonumber
\left[\left(\partial^2_\xi+\frac{1}{\xi}\partial_\xi-\frac{l^2}{\xi^2}\right)+(1+\mu)^2\right]g(\xi)&=&2\eta(1+\mu)\, f(\xi)\,.\\
\eea
For quasi-energies inside the bulk dynamical gap, the wavefunction must decay far from the defect. Hence, let us look for a solution of the form $f(\xi)=c\, K_{l}(\lambda \xi)$ and $g(\xi)=d\, K_{l}(\lambda \xi)$, where $K_l(x)$ is the modified Bessel function of the $2$nd kind that satisfy
\be
\left(\partial^2_\xi+\frac{1}{\xi}\partial_\xi-\frac{l^2}{\xi^2}\right)K_l(\lambda\xi)=\lambda^2K_l(\lambda\xi)\,.
\ee
Introducing this into Eqs. (\ref{eqreduced}) we arrive to the following condition for $\lambda$,
\be
[\lambda^2+(1-\mu)^2][\lambda^2+(1+\mu)^2]=-4\eta^2(1-\mu^2)\,.
\ee
and the relation
\be
\frac{c}{d}=-\frac{2\eta(1-\mu)}{\lambda^2+(1-\mu)^2}\,.
\label{cd}
\ee
The equation for $\lambda$ has four solutions which are complex conjugate in pairs. The two physical solutions correspond to $\mathrm{Re}(\lambda)>0$ as this guarantees  an exponential decay for large $r$. Let us denote these two solutions as $\lambda_+$ and $\lambda_-=\lambda_+^*$, 
\be
\lambda_\pm=\sqrt{-1-\mu^2\pm2\sqrt{-\eta^2+\mu^2(1+\eta^2)}}\,.
\ee
The region where $\mathrm{Re}(\lambda)>0$ corresponds to $|\mu|<\eta/\sqrt{1+\eta^2}$, that is, inside the bulk dynamical gap,\cite{Usaj2014a} $\Delta=\hbar\Omega\eta/\sqrt{1+\eta^2}$.
The other components of the Floquet wavefunction can be readily obtained as
\bea
\nonumber
u_{1A}(\bm{r})&=&\frac{ i  \mathrm{e}^{ i (l-1)\varphi}}{1-\mu}\left(\partial_\xi+\frac{l}{\xi}\right)f(\xi)=\frac{ i  \mathrm{e}^{ i (l-1)\varphi}}{1-\mu}\tilde{f}(\xi),\\
\nonumber
u_{0B}(\bm{r})&=&-\frac{ i \mathrm{e}^{ i (l+1)\varphi}}{1+\mu}\left(\partial_\xi-\frac{l}{\xi}\right)g(\xi)=-\frac{ i \mathrm{e}^{ i (l+1)\varphi}}{1+\mu}\tilde{g}(\xi)\,,\\
\label{eqsol2}
\eea
which are straightforward to evaluate since $\left(\partial_\xi\mp\frac{l}{\xi}\right)K_l(\lambda \xi)=-\lambda K_{l\pm1}(\lambda \xi)$. It is worth to point out that $\la{u_1}|u_0\ra=0$ so that $\phi(\bm{r},t)$ can be normalized for any time $t$ in this approximation,\cite{Usaj2014a} which allows to calculate not only time-averaged quantities but also their time dependence explicitly.\\

To proceed any further we need to specify the defect type, which allows the setting of the appropriate boundary conditions. In the following we present a detailed discussion for some particular but relevant cases. 

\section{Boundary conditions}
\label{BoundaryConditions}
The boundary conditions (BC) must guarantee that the probability current perpendicular to the defect boundary cancels out.
Here, we shall consider only three types of BCs that represent three generic cases and serve to illustrate the overall picture: the zigzag-like BC (ZZBC), the armchair-like BC (ABC) and the infinite mass BC (IMBC).~\cite{Berry1987} 

Since the BC needs to be satisfied at any time, in Floquet space the boundary condition must be imposed on each replica separately. Therefore, the boundary problem is analogous to the static one and  we shall follow Refs. [\onlinecite{McCann2004}] and [\onlinecite{Akhmerov2007}] and use a matrix $\bm{M}$ to  introduce the appropriate relations between the components of the $A$ and $B$ sublattices and the two  Dirac cones at the boundary for the three types of BCs~\cite{Akhmerov2007, Beenakker2008a}.  

An arbitrary BC can be written in  the form 
\begin{eqnarray} 
\label{BC}
 \Psi(r=R(\varphi),\varphi)&=&\bm{M}(\varphi)\Psi (r=R(\varphi),\varphi)\,,
\end{eqnarray}
where $R(\varphi)$ defines the shape of the defect and the matrix $\bm{M}$ (in the isotropic representation) is given by
\be
\bm{M}(\varphi)= (\hat{\bm{\nu}} \cdot \bm{\tau})\otimes(\hat{\bm{n}} \cdot \bm{\sigma})\,.
\label{M-matrix}
\ee
Here $\bm{\sigma}$ refers to the sublattice pseudospin and $\bm{\tau}$ to the valley (Dirac cones) isospin. The matrix
$\bm{M}$ has all the information about the shape of the boundary via the unit vector $\hat{\bm{n}}$. On the other hand, the nature of the honeycomb lattice's termination is related to the unit vector $\hat{\bm{\nu}}$, that rules whether the two Dirac cones mix or not. Namely, for a defect with a straight boundary,~\cite{Akhmerov2007}
\bea
\text{ZZBC}&\to& \hat{\bm{\nu}} =\hat{\bm{z}}\,, \hat{\bm{n}} =\pm\hat{\bm{z}} \nonumber \\
\text{ABC}&\to& \hat{\bm{\nu}} \cdot \hat{\bm{z}} = 0\,, \hat{\bm{n}}=\hat{\bm{z}}\times\hat{\bm{n}}_B \label{BCeqs} \\
\text{IMBC}&\to&\hat{\bm{\nu}}=\hat{\bm{z}}\,, \hat{\bm{n}}=\hat{\bm{z}}\times\hat{\bm{n}}_B\,, \nonumber
\eea 
 where $\hat{\bm{n}}_B$ is an unitary vector perpendicular to the defect boundary and pointing inwards.
 From the above expressions it is clear that while armchair BC mixes cones, zigzag and infinite mass BCs do not.
In the following we shall be interested in the comparison between analytical and numerical results for simple geometries, and so we will restrict ourselves to handle only defects with regular polygonal shapes with $N$ sides. 
The general form of $\bm{M}_N$ for such cases is given in the Appendix \ref{BoundaryConditions_App}.

While for the honeycomb lattice, defects with well defined terminations can only have $N=3$ or $ N=6$, it is useful to discuss the limiting case of a circular defect and then compare with the numerics. For the ABC and IMBC this corresponds to the limit $N\rightarrow\infty$ while for the ZZBC care is needed to account for the change of the sublattice character of the edge atoms [$\hat{\bm{n}}=\pm \hat{\bm{z}}$ depending on the sublattice].

\subsection{Circular defect with ``zigzag" boundary condition\label{ZZBC}}
The ZZBC does not mix valleys. This is valid for arbitrary $N$, i.e. $\bm{M}_N$ is diagonal in the isospin subspace. Moreover, it is also diagonal in the pseudospin subspace. However, it is possible, as in the hexagonal geometry, that different sides of the polygon terminate in sites corresponding to different sublattices. This is represented by the $\hat{\bm{n}}=\pm\hat{\bm{z}}$ in Eq. (\ref{BCeqs}),  where the sign changes from side to side, thereby making it cumbersome to handle analytically. Hence, for the sake of simplicity, we will consider a `fictitious' case where the $\pm$ sign is ignored and later compare with the exact numerical calculation. Hereon we will refer to it as the circular-ZZBC (cZZBC).This will help us to better grasp some aspects of the problem.

\begin{figure}[tb]
\includegraphics[width=0.98\columnwidth]{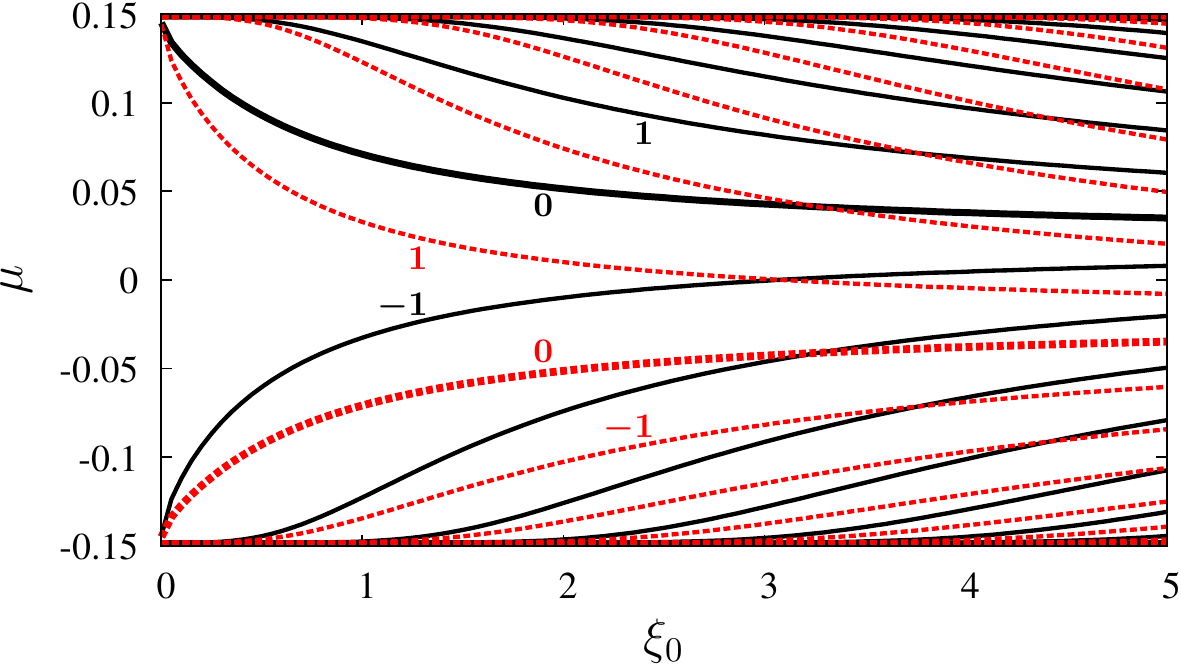}
\caption{(color online) Energy levels for $\eta=0.15$ and $l=0$, $\pm1$, $\pm2$,.. as a function of the size of the defect. Solid and dashed lines correspond to the different Dirac cones. In both cases, the thicker lines correspond to $l=0$ and energy levels with $l>0$ ($l<0$) emerge from the top (bottom) of the dynamical gap.}
\label{nivelesZZBC}
\end{figure}

For a circular defect (of radius $R$) the BC implies, say, that $u_{1B}(|\bm{r}|=R)=0$ and $u_{0B}(|\bm{r}|=R)=0$---this corresponds to a honeycomb lattice that ends on $A$ sites. To satisfy it we need to combine the two independent bulk solutions discussed in Section \ref{model}. That is,
\bea
\nonumber
f_{l}(\xi)&=&c_{+} K_l(\lambda_{+}\xi)+c_{-} K_l(\lambda_{-}\xi)\\
g_{l}(\xi)&=&d_{+} K_l(\lambda_{+}\xi)+d_{-} K_l(\lambda_{-}\xi)\,,
\eea
where we have kept the previous notation. Then we have that  
\be \label{cond_zzbc}
f_{l}(\xi_0) = 0, \qquad \tilde{g}_{l}(\xi_0) = 0,
\ee
with $\xi_0=k_0R$. This leads to  the following relations between coefficients: $|c_{+}|=|c_{-}|$ and $|d_{+}|=|d_{-}|$. 
By introducing them back into Eqs. (\ref{eqreduced}) we obtain, for the $K$ cone, the following equation for the quasi-energy ($\mu$) 
\be
\mathrm{Im}[\beta_+\lambda_+ K_{l}(\lambda_-\xi_0)K_{l+1}(\lambda_+\xi_0) ]=0\,. 
\label{condKcone}
\ee
with
\be
\beta_\pm = -\frac{\lambda_\pm^2 + (1-\mu)^2}{2\eta (1-\mu)}\,.
\ee
The solutions ($\mu_l$) to this equation  form a discrete set of quasi-energies inside the bulk dynamical gap. 
Figure \ref{nivelesZZBC} shows them as a function of $\xi_0$ (throughout this work, we shall use $\eta = 0.15$ and $\hbar \Omega = 0.1 \, t$ in all numerical calculations).
Notice that the symmetry between $l>0$ and $l<0$ is broken by the radiation field.

The symmetry of the Floquet spectrum around the center of the gap ($\mu=0$) is recovered when the com\-ple\-men\-tary valley ($K'$ cone) is considered. For that, we recall that the solutions for the $K'$ cone can be obtained by relabeling the Floquet wavefunction as $\phi'(\bm{r})=\{[-u'_{1B}(\bm{r}),u'_{1A}(\bm{r})],[-u'_{0B}(\bm{r}),u'_{0A}(\bm{r})]\}^\mathrm{T}$ (see the appendix). This results in an additional  set of quasi-energies that can be obtained from the condition 
\be
\mathrm{Im}[\beta_- \lambda_+ K_l(\lambda_- \xi_0)K_{l-1}(\lambda_+ \xi_0)] = 0. \label{condKpcone}
\ee
It can be shown that the latter set of quasi-energies can be obtained from 
Eq. (\ref{condKcone}) by exchanging $(l,\mu)\rightarrow (-l,-\mu)$, which is precisely what is needed to recover the symmetry around 
$\mu=0$.

It is interesting to consider, for a fixed $l$, the limit of very large radii, $\xi_0 \gg \xi_\mathrm{d} = k_0\, \hbar v_F/\Delta = \sqrt{1+\eta^2}/2\eta$ and  approximate $K_{l}(\lambda \xi_0)$ by its asymptotic expansion. By doing so, Eqs. (\ref{condKcone}) and Eq. (\ref{condKpcone}) leads to 
\be
\mu_l = \pm \eta^2+\frac{(l\pm1/2)\eta}{\xi}+\mathcal{O}(\xi^{-2},\eta^2)\,,
\ee
respectively. This result can be understood in terms of the quasi-energy dispersion of the edge states in irradiated semi-infinite graphene sheets with a zigzag termination.\cite{Usaj2014a} In that case, it was shown that, close to the center of the gap, the quasi-energy dispersion  can be approximated by $\varepsilon_k=\hbar\Omega/2\pm \hbar\Omega\eta^2/2+\hbar v_F\eta k$. Our result for $\mu_l$ is then reflecting the fact that the wavevector $k$ along the defect's edge must be quantized, 
\be
k_l= \frac{(l\pm 1/2)}{R}\,.
\ee
 It is worth mentioning that in this large radii limit the Floquet states have roughly the same weigth on the two Floquet replicas. 

\begin{figure}[t]
\includegraphics[width=0.98\columnwidth]{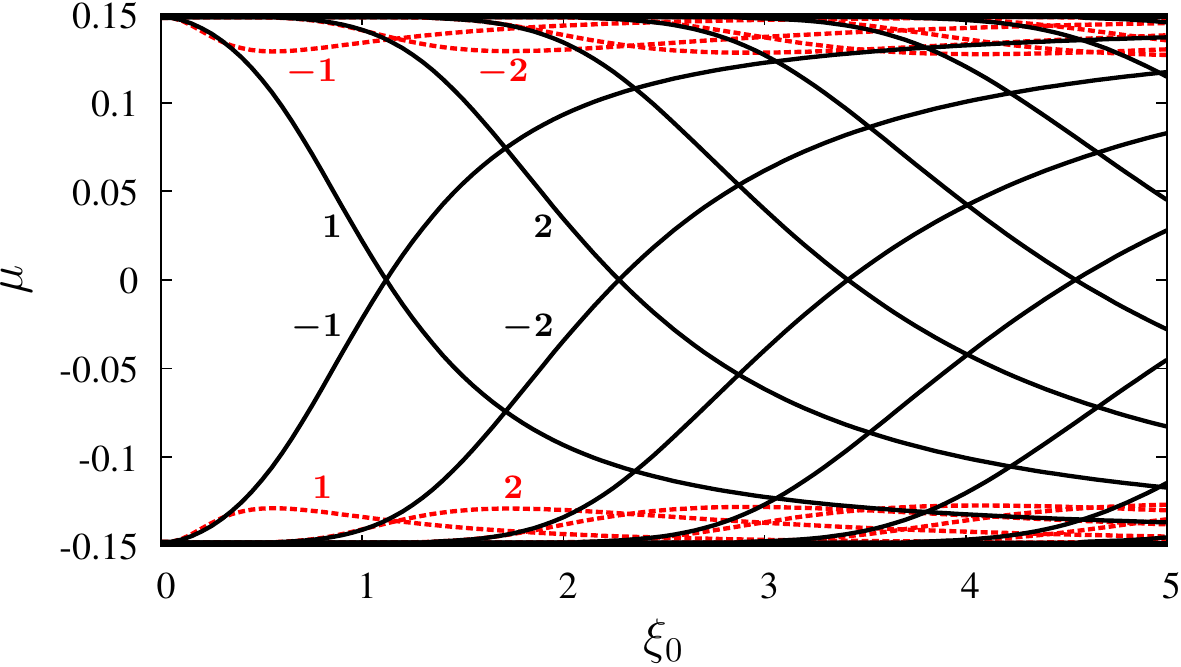}
\caption{(color online) Energy levels for the case of IMBC. Parameters as in Fig. \ref{nivelesZZBC}. The solid (dashed) line corresponds to $K(K')$ cone. There is no solution with $l=0$.}
\label{nivelesIMBC}
\end{figure}

\subsection{Infinite mass boundary condition}
\label{IMBC}
The IMBC was introduced by Berry and Mondragon in Ref. [\onlinecite{Berry1987}] to study confined Dirac particles (`neutrino billiards').
It corresponds to add a mass term to the Dirac equation only in a given region of space (in our case the defect) and take the limit of such a mass going to infinity. While this could be thought as a local staggered potential in the honeycomb lattice, it must be kept in mind that this is only the case for a staggered potential much smaller than the bandwidth--this is so because if the  staggered potential is too large it behaves like an effective hole (introducing inter-valley scattering depending on the geometry of the defect).  The latter limit was not a problem in Ref. [\onlinecite{Berry1987}] , because they only considered a single unbound massless Dirac particle.

Since the  IMBC does not mix valleys either, we can treat again both Dirac cones separately. We start by using the circular geometry, which corresponds to the $N\to\infty$ limit of $\bm{M}_N$. For the IMBC  $\bm{M}_{\infty}$ is not longer diagonal in the pseudospin subspace and thus the $A$ and $B$ components of the wavefunction are not independent any more. In fact, Eq. (\ref{BC}) requires that~\cite{Berry1987} 
\be
\frac{u_{jB}(R,\varphi)}{u_{jA}(R,\varphi)}=-ie^{i\varphi}\,,\qquad \frac{u'_{jB}(R,\varphi)}{u'_{jA}(R,\varphi)}=ie^{-i\varphi}\,,
\ee
for the $K$ and $K'$ cone, respectively, where $j=0,1$ is the Floquet subspace index ---notice that $\lim_{N \to \infty} \Xi_N (\varphi) = i e^{-i\varphi}$ in the definition of the $\bm{M}_N$ matrix, see appendix.
Following the same procedure as in the previous section, and using the same notation,  these conditions imply that 
\bea
\nonumber
(1-\mu)\,f_{l}(\xi_0)&=&\pm \tilde{f}_{l}(\xi_0)\,,\\
 (1+\mu)\,g_{l}(\xi_0)&=&\pm\tilde{g}_{l}(\xi_0)\,,
\eea
while the equation for the  quasi-energies is given by 
\begin{widetext}
\be
\mathrm{Im}\left[\left(\lambda_+^2+(1-\mu)^2\right)\left\{\lambda_+ K_{l+1}(\lambda_+\xi_0)\mp(1+\mu)K_l(\lambda_+\xi_0)\right\}\left\{\lambda_- K_{l-1}(\lambda_-\xi_0)\mp(1-\mu)K_l(\lambda_-\xi_0)\right\}\right]=0.
\label{eqIMBC}
\ee
\end{widetext}
Here the (-) and (+) signs correspond to $K$ and $K'$ cone, respectively. 
It can be shown that the above expression remains invariant under the change $(\mu, l)\rightarrow (-\mu,-\l)$ for each cone separately and, therefore, unlike the cZZBC, the Floquet spectrum for the IMBC is symmetric around  $\mu=0$ for each cone. Using this symmetry of Eq. (\ref{eqIMBC}) it is straightforward to verify that there is no solution for $l=0$ (that necessarily corresponds to $\mu=0$). The  IMBC Floquet spectrum  is shown in Fig. \ref{nivelesIMBC} as a function of $\xi_0$ . 
Note that the two cones have a completely different spectrum. This could be anticipated from the fact that the presence of both the staggered potential and the radiation field breaks the valley symmetry (cf. Fig. \ref{cones} below)---it is worth mentioning that the bulk Floquet gap at $k=0$ can even present a topological phase transition depending on the relative magnitude of the mass term and the radiation field.~\cite{Ezawa2013}

When defects are made of regular polygons, i.e. with finite $N$, the $\bm{M}_N$ matrix acquire a non-trivial structure as a function of $\varphi$. Thus, the states whose quantum numbers $l$ differ in $N$ are coupled, thereby leading to avoided crossings. The equations for this case are rather cumbersome (some of them are presented in the appendix) but can be solved in a perturbative fashion. 
Some examples are presented in Sec \ref{TBM} in comparison with the numerical solutions of the tight-binding model.

\subsection{Armchair boundary condition}
The ACB is analog to the IMBC in the pseudospin subspace, leading to similar quasi-energy spectra. The difference between both boundary conditions rely on the isospin subspace: while ACB mixes cones, IMBC does not. Thus, ACB exhibits additional avoided crossings between modes belonging to different cones (see numerical results in Sec. \ref{TBM}). Because cones are mixed, they both need to be treated together and hence the dimension of the Floquet space is doubled. The analytical procedure is similar to the one presented for the other BCs, whose details are beyond the scope of the present work. We will then limit, for this case, to discuss the numerical results in in Sec. \ref{TBM}.
\begin{figure}[t]
\includegraphics[width=0.95\columnwidth]{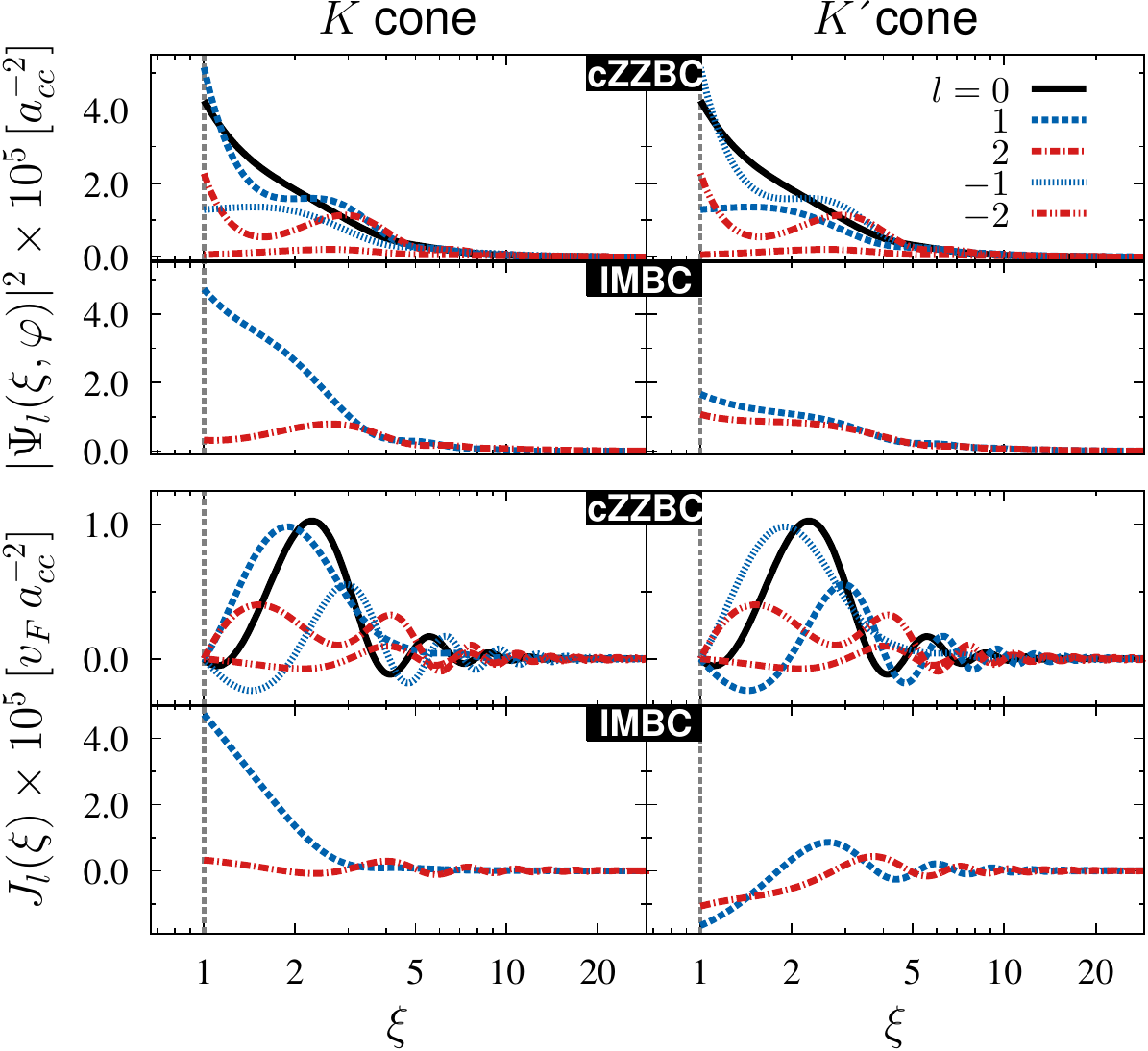}
\caption{(color online) Probability (top) and current (bottom)  densities  as a function of the radial coordinate $\xi$ for two different boundary conditions studied in Sec. \ref{BoundaryConditions}. Notice the log scale on the horizontal axis. In all cases, the defect boundary is located at $\xi_0=1$ ($R=30\,a_{cc}$). 
Probabilities and current densities with $|l|>3$ are several orders of magnitude smaller than the showed in the figure and was omitted for clarity. In the case of the IMBC, curves with $l$ and $-l$ are coincident ($l>0$ are showed).
The spatial range showed in the figure correspond to the distance from the centre of the defect to the end of the samples of graphene sheets used in Sec. \ref{TBM} for the numerical tight-binding calculations $(500\times\sqrt{3}\,a_{cc})$.}
 \label{densities}
\end{figure}

\section{Probability current density: chiral current}
\label{chiral_current}
So far we have mainly analyzed the spectrum of the Floquet bound states inside the dynamical gap (around $\hbar\Omega/2$) for a circular defect. Now we focus on their chiral nature. The velocity operator is given by $\hat{\bm{v}}=v_F \bm{\sigma}$ and hence the time averaged (over one period) probability current density is
\begin{eqnarray}
\nonumber
\bm{J}(\bm{r})&=&v_F\,\phi^\dagger(\bm{r})\bm{\sigma}\phi(\bm{r})\\
&=&\left(\la \sigma_r\ra_1+\la \sigma_r\ra_0\right)\,\hat{\bm{r}}+\left(\la \sigma_\varphi\ra_1+\la \sigma_\varphi\ra_0\right)\,\hat{\bm{\varphi}}\,,
\end{eqnarray}
where $\la \sigma_\alpha\ra_j=\{u_{jA,l}^*(\bm{r}),u_{jB,l}^*(\bm{r})\}\sigma_\alpha\{u_{jA,l}(\bm{r}),u_{jB,l}(\bm{r})\}^T$, $j=0,1$ is the same as earlier, $\sigma_r=\bm{\sigma}\cdot\hat{\bm{r}}$ and $\sigma_\varphi=\bm{\sigma}\cdot\hat{\bm{\varphi}}$.
Using the solutions founded in the previous section, it can be readily shown that
\bea
\nonumber
\la \sigma_r\ra_1&=&-\frac{2}{1-\mu_{l}}\mathrm{Im}\left(f_{l}(\xi)\tilde{f}_{l}^*(\xi)\right)\\
\nonumber
\la \sigma_r\ra_0&=&-\frac{2}{1+\mu_{l}}\mathrm{Im}\left(g_{l}(\xi)\tilde{g}_{l}^*(\xi)\right)\\
\nonumber
\la \sigma_\varphi\ra_1&=&-\frac{2}{1-\mu_{l}}\mathrm{Re}\left(f_{l}(\xi)\tilde{f}_{l}^*(\xi)\right)\\
\la \sigma_\varphi\ra_0&=&-\frac{2}{1+\mu_{l}}\mathrm{Re}\left(g_{l}(\xi)\tilde{g}_{l}^*(\xi)\right)\,.
\eea
Since $\lambda_+=\lambda_-^*$, one can easily check that $\mathrm{Im}\left(f_{l}(\xi)\tilde{f}_{l}^*(\xi)\right)=\mathrm{Im}\left(g_{l}(\xi)\tilde{g}_{l}^*(\xi)\right)=0$ so that the radial component of the current density vanishes, as expected. Therefore, we have
\be
\bm{J}_{l}(\xi)=-2\, v_F\,\left(\frac{f_{l}(\xi)\tilde{f}_{l}^*(\xi)}{1-\mu_{l}}+\frac{g_{l}(\xi)\tilde{g}_{l}^*(\xi)}{1+\mu_{l}}\right)\,\hat{\bm{\varphi}}\,.
\ee

\begin{figure}[t]
\includegraphics[width=0.95\columnwidth]{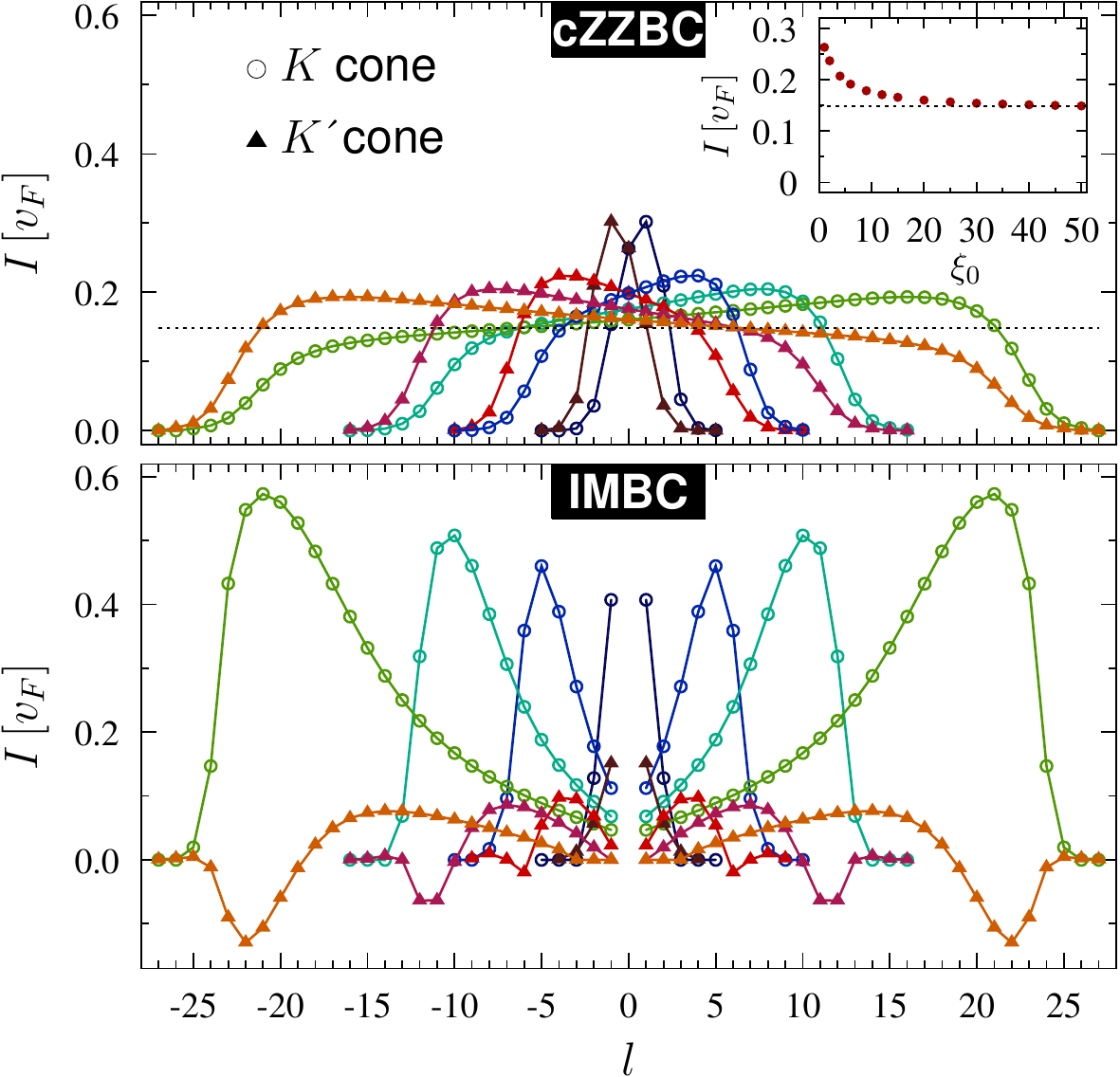}
\caption{(color online) Total current as a function of the  quantum number $l$ for different defect sizes ($\xi_0 = 1,5,10,20$). Open (close) symbols  correspond to Floquet states lying in the $K$ ($K'$) cone (lines are only guides for the eye). A curve with a larger span on $l$ corresponds to a larger $\xi_0$. The dotted black line in the top panel represents the $\eta/(1+\eta^2)$ value expected for the current of the Floquet edge state in a semi-infinite irradiated graphene sheet with zigzag termination. The inset shows how this limit is reached for $l=0$ when the size of the defect increases.}
 \label{currents}
\end{figure}

Figure \ref{densities} shows the spatial dependence of both the probability  and the current density for the $K$ and $K'$ cones and for the two different boundary conditions analyzed in Sec. \ref{BoundaryConditions}. The curves  correspond to a defect of $R=30\, a_{cc}$, i.e., $\xi_0=1$ with the parameters used throughout this work. 
We have only retained the Floquet wavefunctions with $|l|=0,1,2$, whose corresponding quasi-energies can be seen from Fig. \ref{nivelesZZBC} and Fig. \ref{nivelesIMBC} for $\xi_0=1$. 
Due to the oscillating nature of the Floquet wavefunctions both probability density functions and current densities show relative maxima and minima (with the same or different signs in the case of current densities) as a function of $\xi$. 
Nevertheless, all of them decay exponentially away from the edge of the defect. This is more evident for the Floquet wavefunctions whose quasi-energies are close to the middle of the dynamical gap as in that case the decay length is shorter. For  quasi-energies close to the edges of the dynamical gap, the decay length becomes larger and larger and the $\xi^{-1/2}$ power law decay, characteristic of the $K_l$ Bessel functions with purely imaginary argument becomes apparent. In these latter cases, however, the current amplitude becomes several orders of magnitude smaller than in the formers (see Fig. \ref{currents}).
For the cZZBC, Fig. \ref{densities} shows the equivalent role that play the $K$ and $K'$ cones under the change $l\leftrightarrow -l$, as it was explained before in Sec. \ref{ZZBC}.
Unlike the cZZBC, for the IMBC the $K$ and $K'$ cones are inequivalent. In this case, as discussed in Sec. \ref{IMBC}, the change $l\leftrightarrow -l$ lead to the same probability and current densities for each cone separately.

\begin{figure}[t]
\hspace{-0.1cm}\includegraphics[width=0.95\columnwidth,height=0.55\columnwidth]{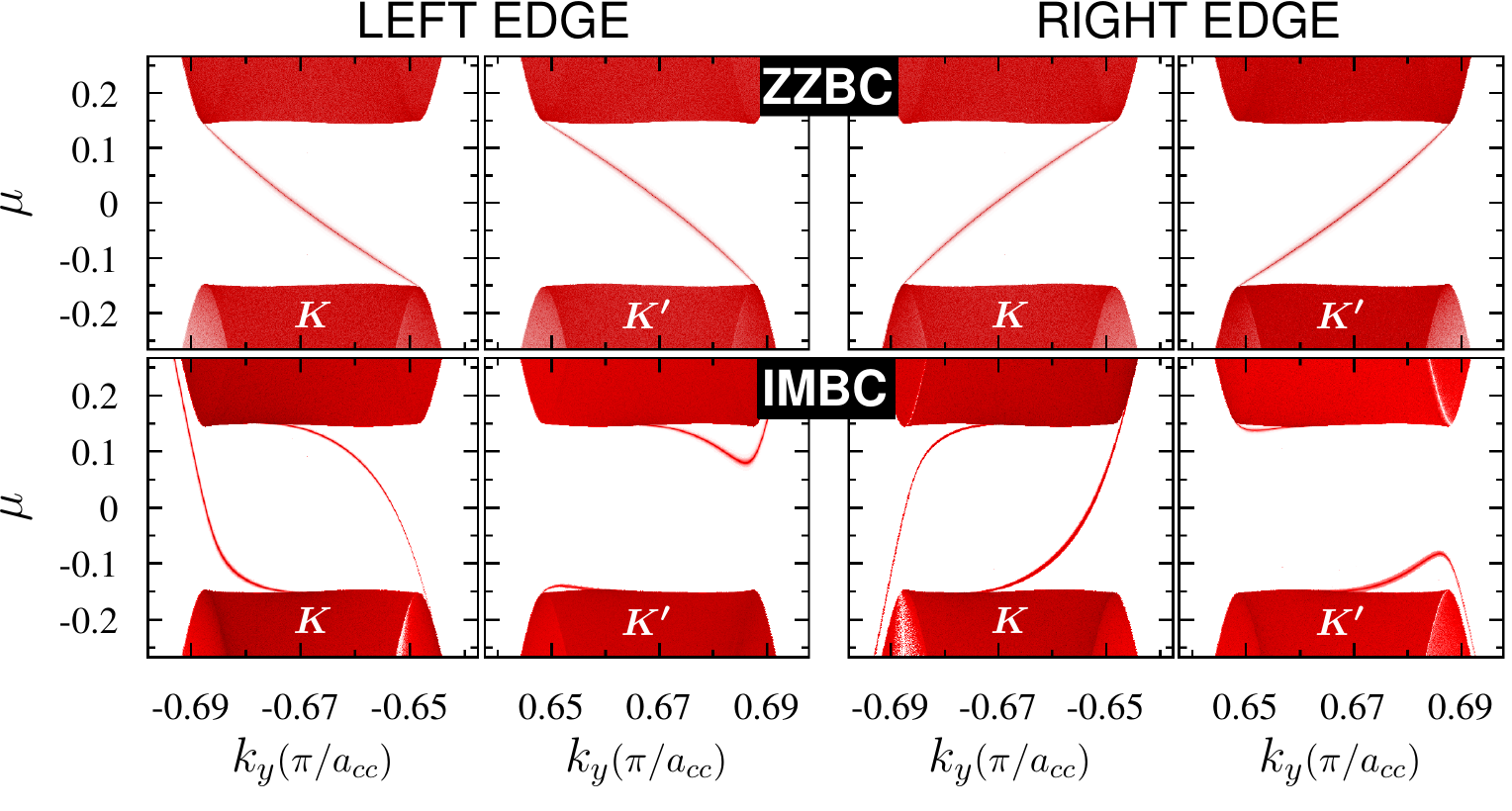}
\caption{(color online) Dispersion relation of a nanoribbon of $\sim 10^6$ atoms width with zigzag (top row) and infinite mass (down row) boundary conditions obtained numerically by decimation procedures and a tight binding model. All parameters like $\hbar\Omega$, $\eta$ and mass $\delta$ are the same used in Sec. \ref{TBM}.}
 \label{cones}
\end{figure}

The lack of equivalence between the $K$ and $K'$ cones for defects with IMBC is also present in systems other than circular defects. For illustrating purposes, Fig. \ref{cones} shows the $k$-dependent local density of states (LDOS) for a nanoribbon with both cZZBC and IMBC, projected on the  $m=0$ Floquet replica. 
Notice that, unlike the cZZBC, the IMBC presents an asymmetry (at each edge) with respect to the middle of the dynamical gap. The symmetry is broken by the presence of the mass term at the edges and it is only globally recovered when both edges are considered---this is so because for zigzag nanoribbons, as considered here, the atoms at the two edges belong to different sublattices.

Even when the current density oscillates as it decays away from the defect, the total current (current densities integrated on $r$) for cZZBC has the same sign for all the bound states.
This is the signature of the chirality of the Floquet states and their signs only depends on the sign of the helicity of the circularly polarized radiation field. Figure \ref{currents} shows the total currents for both cZZBC and IMBC as a function of the  quantum number $l$ for defects with $\xi_0 = 1,5,10,20$. 
Unlike the cZZBC, the IMBC only presents chiral Floquet states for the $K$ cone. 
Analogously, Fig. \ref{cones} shows a similar behavior for the nanoribbon with IMBC: while the $K$ cone presents two chiral states at each edge, $K'$ cone has none. 

Finally, it is interesting to analyze the value of the total current of a given bound state in the limit of a large defect. As discussed in Sec. \ref{ZZBC} for large $R$ the quasi-energy dispersion can be related to the one corresponding to a nanoribbon as the boundary of the defect appears (locally) as a straight line (i.e. when the radius is much larger than the decay length). In that case the expected velocity for each bound states is $v=\hbar^{-1}\partial \varepsilon_k/\partial k\simeq v_F\eta$, or more precisely $v=v_F\eta/(1+\eta^2)$.~\cite{Usaj2014a}
 The inset of the Fig. \ref{currents} shows the current in units of $v_F$ for Floquet states with $l=0$ (red points) as a function of the size of the defects. The black dotted line represent the expected $\eta/(1+\eta^2)$---this is also indicated in the main figure. Clearly, there is a good agreement with the expected value. A similar behavior is observed for states with different quantum number $l$ as the size of the defect increases.

\section{Comparison with the Tight-binding model\label{TBM}}
In this section, we calculate the  quasi-energy spectra within the dynamical gap numerically as a function of the size and shape of the defect for all three types of boundary conditions mentioned before, ZZBC, ABC and IMBC, and compare with the analytical results when possible.

In order to describe the electronic structure of irradiated graphene sheets near the Fermi energy, we resort to the widely used tight-binding Hamiltonian,~\cite{Wallace1947,Saito1998,Charlier2007} which is  written only in terms of $p_z$ orbitals with energies $\epsilon_i$ for a given carbon atom located at site $i$ and hopping matrix elements $\gamma_{ij}$ between nearest-neighbors carbon atoms. In second quantization notation, it results
\be
	\mathcal{H} = \sum_{i} \epsilon_i\, c_i^{\dagger} c_i - \sum_{\langle i,j \rangle} \left( \gamma_{ij} \, c_i^{\dagger} c_j + \mathrm{H.c.} \right),
\ee
where the operator $c_i^{\dagger}(c_i^{})$ creates (annihilates) a $p_z$-electron on site $i$. 
The effect of the laser is introduced  through the  time-dependent phase of the hopping matrix elements,\cite{Oka2009,Calvo2012a,Calvo2013}
\begin{equation}
\gamma_{ij}=
\gamma_{0}\exp\left(\mathrm{i}\frac{2\pi}{\Phi_0}\int_{
\bm{r}_i}^{\bm{r}_j}\bm{A}(t)\cdot\mathrm{d}\bm{\ell}\right)\,,
\label{gama}
\end{equation}
where $\Phi_0$ is the magnetic flux quantum and $\gamma_0 \sim2.7$ eV~\cite{Dubois2009}. 

By using Floquet theory \cite{Kohler2005,Platero2004,Moskalets2002} as described before one can compute the Floquet spectrum. Once again, one ends up with a time-independent problem in an expanded space. In this case one can picture it as tight-binding problem in a multichannel system where each channel represents the graphene sheet with different number of photons.\cite{Shirley1965, Calvo2013, FoaTorres2014a}  It is worth mentioning that in the tight-binding method the time dependent perturbation is never purely harmonic given the exponential dependence of Eq. (\ref{gama}) on the radiation field amplitude. Hence, there is a coupling among all the replicas\cite{Calvo2013} and not just those with $\Delta m=\pm1$. Nevertheless, for $\eta\ll 1$, only the latter are relevant. 

Because the problem in the Floquet space becomes time independent, one can use standard techniques to calculate the quasi-energy spectrum. In this case we used the Chebyshev's polynomials method~\cite{Weise2006} which provides an order $N$ method of proven efficiency~\cite{Lherbier2008}. This allows us to tackle very large systems sizes so that our defect is far from the boundaries and can be considered as a 'bulk defect'. For simplicity we only retained two Floquet replicas just like its theoretical counterpart studied in Section \ref{model}. This is a good approximation whenever $\eta\ll1$. The addition of more replicas would lead to the development of a hierarchy of bound states in a similar way as for edge states at the border of an irradiated graphene sample.\cite{Perez-Piskunow2015}

Defects were introduced in graphene by defining geometrical shapes---triangles, hexagons, and circles---and removing all atoms inside it (for the ZZBC and ABC) as well as any remaining dangling bonds. In the case of the IMBC,  a staggered potential was introduced only inside the defect---i.e. we added on-site energies ($\pm \delta$)  whose signs depend on the sublattice index. In all calculations we used $\delta=\gamma_0/2$, which is larger than $\hbar\Omega/2$ (taken to be $\sim\gamma_0/20$) but not too large as to become equivalent to a hole ($\delta\to \infty$ is equivalent to a hole defect).
Triangles and hexagons in arbitrary orientations lead to edges with mixed zigzag and armchair terminations. However, for specific orientations with respect to the C-C bonds, it is possible to construct defects with only one termination type---we will refer to them as zigzag/armchair triangular and hexagonal defects. 
Circles, of course, are always a mixture of different edge terminations and, as we will show, present some special features.
In all cases, the numerical calculations were performed using graphene samples of $1000 \times 1000$ unit cells. 

\begin{figure}[tb]
\includegraphics[width=0.99\columnwidth]{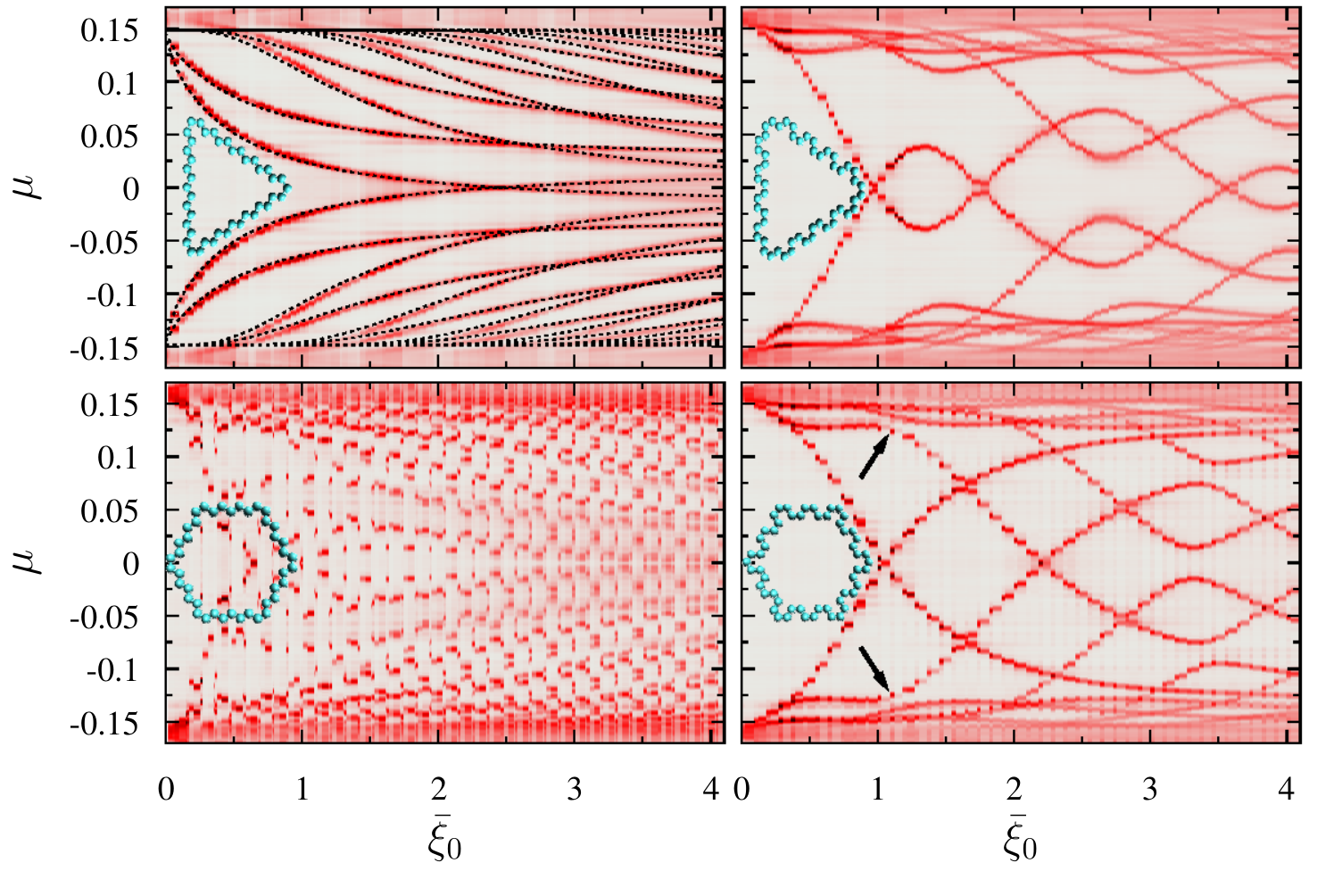}
\caption{(color online) Color map of the Floquet local density of states (FLDOS) projected on the $m=0$ replica and onto sites located around the boundary of the defects  for different sizes of the defects, $\bar{\xi}_0=k_0\bar{R}$ (see main text). Top and bottom panels show the case of triangular and hexagonal holes, respectively, with zigzag (left ) and armchair (right) edge termination. The appearance of Floquet bound states inside the bulk dynamical gap is apparent from the figure. The dashed lines in the zigzag triangular case correspond to the analytical solution found in Sec. \ref{ZZBC} for a `zigzag circle'.}
 \label{TH-hole}
\end{figure}

\begin{figure}[t]
\includegraphics[width=0.99\columnwidth]{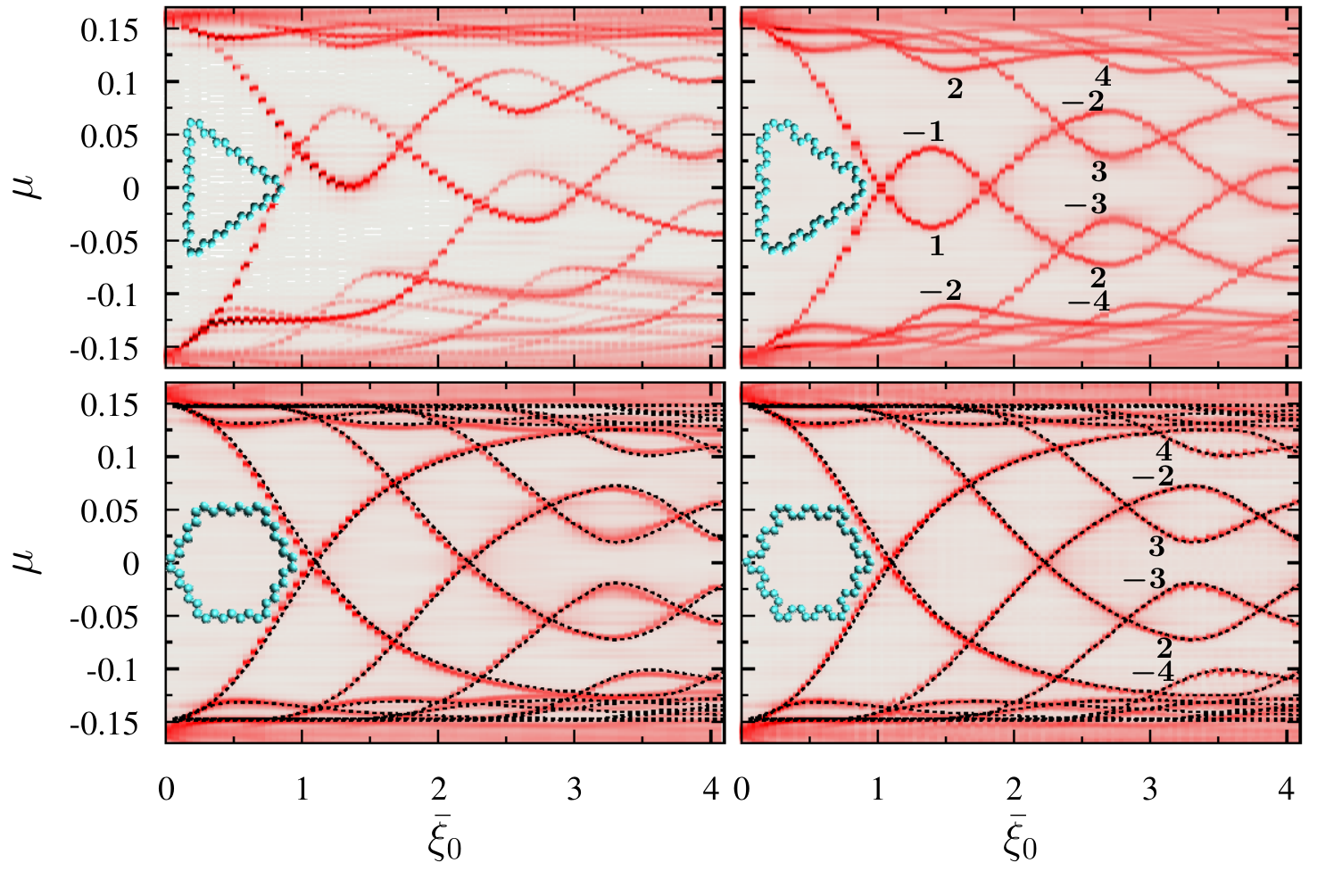}
\caption{(color online) Same as Fig. \ref{TH-hole} but for the case of where a staggered potential is included inside the defect region (IMBC). Dashed lines on the bottom panels correspond to the analytical solutions calculated for the IMBC hexagon as explained in the appendix. Notice that, unlike the hole defects, the FLDOS for the IMBC does not depend on the termination of the defects---except for the zigzag triangular defect (see main text). }
\label{TH-im}
\end{figure}

Figures \ref{TH-hole} and \ref{TH-im} show a color map of the Floquet local density of states (FLDOS) inside the bulk gap  (projected onto a few sites around the defect boundary, and on the $m=0$ replica) as a function of the size of the defect for hole and staggered potential defects, respectively.  The shape of the defect is indicated in the figures. Left panels correspond to zigzag terminations and the right panels to the armchair ones. Dashed (black) lines correspond to the solutions obtained from the continuum model (see discussion below).
It is apparent from the figures that discrete Floquet bound states do appear inside the dynamical gap. Interestingly, in most cases,  the quasi-energy spectrum resemble the ones obtained with the analytical model proposed in Sec. \ref{model}.
This remains valid for the triangular shaped zigzag hole even when the analytical solution relies on the circular symmetry of the defects.  
It is worth mentioning that for a quantitative comparison an effective radius is needed. In these cases we used $\bar{R}=1/(2\pi)\int_0^{2\pi} R(\varphi)\, d\varphi=R_0a_{0,N}$ (see appendix).
%

There are few points worth to emphasize:
\begin{itemize}
\item[(i)] avoided crossings are observed in most cases due to the discrete rotational symmetry of the defect that introduces a $\varphi$ dependence on $\bm{M}$, as well as of the boundary radius $R(\varphi)$, as discussed in Sec. \ref{BoundaryConditions} and the appendix.  This avoided crossings occurs whenever the quantum numbers of the crossing levels,  $l$ and $l'$, differ in a multiple of the number of sides $N$. A few particular examples are indicated in the Fig. \ref{TH-im}.
\item[(ii)] the latter picture  is very particular in the case of the zigzag triangular hole defect (top-left in the Fig. \ref{TH-hole}). On the one hand,  the matrix $\bm{M}$ is independent of $\varphi$---note that $\hat{\bm{n}}=\hat{\bm{z}}$ for any $\varphi$ as the edge site always belong to the same sublattice and the direction of $\hat{\bm{\nu}}$ is fixed for each cone---and hence the only dependence on $\varphi$ appears through the boundary radius $R(\varphi)$. On the other hand, for each cone, the `unperturbed' energy levels of the `zigzag circle' are never degenerated, making the effect even weaker. As a result, the energy level are well described by assuming that there is no mixing between states with different quantum number $l$.  Notice also there is no mixing between different cones or valleys.
\item[(iii)] the zigzag triangular defect with the staggered potential shows a shift in energy with respect to the IMBC solution. This is related to the sublattice imbalance of the edge sites and the fact that both sublattices have different energy inside the defect (staggered potential). This effect is not observed for the other geometries as they have balanced edges.
\item[(iv)] the armchair hexagonal hole defect shows two distinct contributions to the quasi-energy spectrum. The one shown in Fig. \ref{TH-hole}, that is very close to the analytical solution for IMBC [except for the anticrossings between energy levels belonging to different cones that are only present in the armchair case (black arrows)], and the one presented in Fig. \ref{H-hole_append} of the appendix, that follow a completely different pattern. The two cases differ in the way the atom chains that constitute each side match at the vertices.
\item[(v)] the zigzag hexagonal hole defect presents a rather complex spectrum quite different from the rest. This is related to the strong mixing between states with different $l$ imposed by the BC that requires that alternating components of the wavefunction cancel in alternating sides.  A precise description of this case is beyond the scope of the present work.
\end{itemize}

\begin{figure}[t]
\includegraphics[width=0.8\columnwidth]{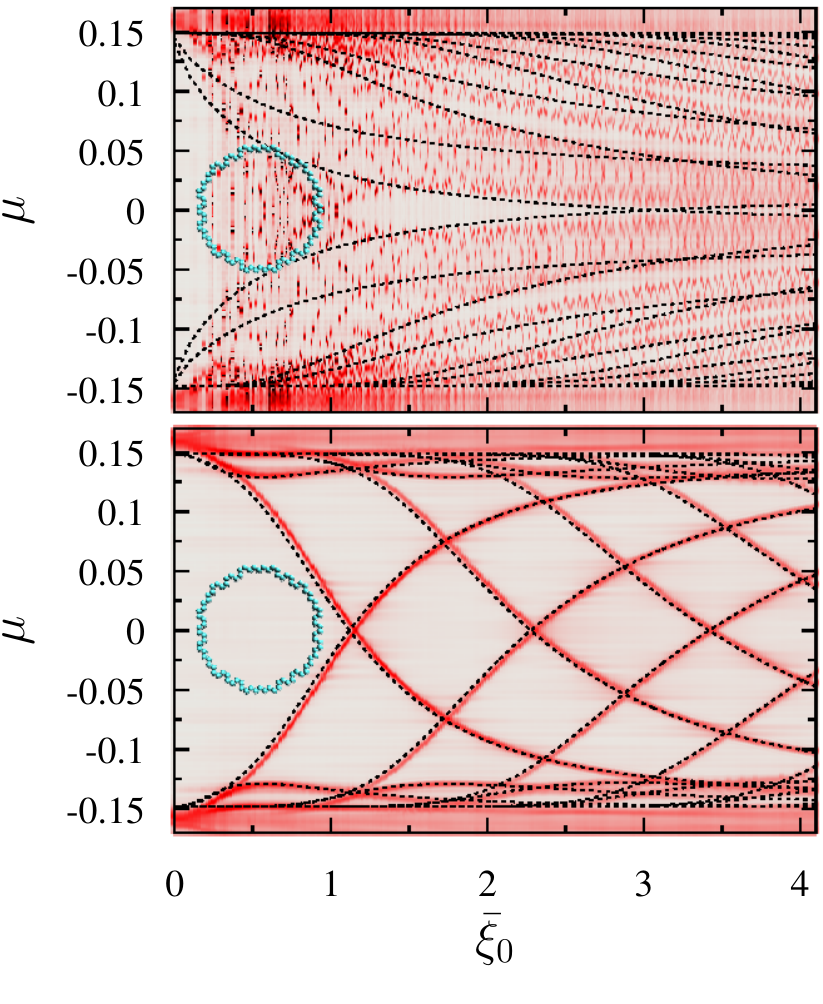}
\caption{(color online) Same as Fig. \ref{TH-hole} but for a circular defect: (i) hole (top); (ii) staggered potential (bottom).}
\label{circles}
\end{figure}

Finally, we show numerical results for circular defects in the Fig. \ref{circles}. The top panel corresponds to a hole defect and the bottom one to the staggered potential defect. Clearly, the latter is very well described by the analytical solutions (dashed black lines). 
 Notice that no avoided crossings (if they exist) are resolved in our numeric simulations, presumably because they are very small since the actual geometry of the  defect is very close to a circle.
The spectrum of the circular hole defect is, as in the zigzag hexagonal one, very complex. Here, however, a more regular pattern emerges for large $R$ as the quasi-energy of the bound states are pretty much confined to regions delimited by the analytical solution of the zigzag circular defect (dashed lines).

One of the questions that remains is to what extent do these bound states survive in the limit of a vacancy defect or, more generally, in the case of adatoms. This is particularly important as the presence of bound states around such impurities might hinder the ability to resolve the laser-induced gaps in actual experiments or lead to percolating states in dirty samples. 

\section{The adatom and vacancy defects\label{adatom_sec}}
The continuum model presented in Sec. \ref{model} is not adequate for analyzing the vacancy limit. In fact, in the $R\to0$ limit for zigzag hole (the appropriate one for a vacancy defect) one finds that there are no solutions inside the gap. Of course, this is not the correct approach as one should introduce a spatial cutoff to account for the finite size of the defect. In this sense, a tight-binding model approach is more convenient and allows for its generalization to include the adatom case.

Since we focus on the bound states within the dynamical gap at $\hbar\Omega/2$, it is enough to consider, as before, only two Floquet replicas, $m=0$ and $m=1$. While for the numerical calculations we will use the real space version of the tight-binding Hamiltonian presented in the previous section, for the discussion of the main aspects of the problem it is better to use a $\bm{k}$-space representation. Then, the Floquet Hamiltonian is written as  
\begin{eqnarray}
\nonumber
\tilde{\mathcal{H}}_F&=&\sum_{\bm{k}}\hbar\Omega(a^{\dagger}_{1\bm{k}}a_{1\bm{k}}^{}+b^{\dagger}_{1\bm{k}}b_{1\bm{k}}^{})\\
&&-t\sum_{\bm{k},m=0,1}(\phi_{\bm{k}}\,a^{\dagger}_{m\bm{k}}b_{m\bm{k}}^{}+\phi^{*}_{\bm{k}}\,b^{\dagger}_{m\bm{k}}a_{m\bm{k}}^{})\\
\nonumber
&& +\sum_{\bm{k}}[A_{\bm{k}}(a^{\dagger}_{1\bm{k}}b_{0\bm{k}}^{}+b^{\dagger}_{1\bm{k}}a_{0\bm{k}}^{})+A_{\bm{k}}^{*}(b^{\dagger}_{0\bm{k}}a_{1\bm{k}}^{}+a^{\dagger}_{0\bm{k}}b_{1\bm{k}}^{})]\,.
\end{eqnarray}
Here $a^{\dagger}_{m\bm{k}}$ and $b^{\dagger}_{m\bm{k}}$ create an electron on the Floquet replica $m$ on the Bloch state with momentum  $\bm{k}$ on the sublattice $A$ and $B$, respectively,  $\phi_{\bm{k}}=\sum_{\bm{\delta}_j} e^{j\bm{k}\cdot\bm{\delta}_j}$, where 
$\{\bm{\delta}_i\}$ are the relative coordinates of the three nearest neighbors $A$ sites of a given $B$ site,  $t=\gamma_0 J_0(z)$ , and  $A_{\bm{k}}=\gamma_0 J_1(z)\sum_{\bm{\delta}_j} e^{i\bm{k}\cdot\bm{\delta}_j}(\bm{\delta}_{jx}-i\bm{\delta}_{jy})/a_{cc}$  with $J_n(x)$ the $n$-th Bessel function of the first kind and $z=2\pi A_0a_{cc}/\Phi_0$.\cite{Calvo2013}

We describe the adatom impurity with a single orbital of energy $\epsilon$ bounded to the C atom at the origin. 
The Hamiltonian of the impurity in the Floquet representation is
 \begin{equation}
\mathcal{H}_{\mathrm{imp}}=\epsilon\, f^{\dagger}_{0}f_{0}^{}+(\epsilon+\hbar\Omega) f^{\dagger}_{1}f_{1}^{}\,,
\end{equation}
and the hybridization term is
 \begin{equation}
\mathcal{H}_{\mathrm{hyb}}=\sum_{\bm{k},m=0,1}V[ f^{\dagger}_{m}a_{m\bm{k}}^{}+a^{\dagger}_{m\bm{k}}f_{m}^{}]\,.
\end{equation}
Note that the the coupling matrix element $V$ does not depend on the radiation field as we are considering normal incidence, hence the phase factor appearing in Eq. (\ref{gama}) is zero. The vacancy limit can be obtained from here by taking $V\to\infty$.

We define the Green function matrix $\bm{\mathcal{G}}$ with elements given by $\mathcal{G}_{ij}=\langle\langle f_{i}^{},f_{j}^{\dag }\rangle\rangle$. Using the Dyson equation it can be written as 
\begin{equation}
 \bm{\mathcal{G}}(\omega)= \left( \begin{array}{ccc}
\omega-\hbar\Omega-\epsilon-V^{2} G_{11}(\omega)& -V^{2} G_{10}(\omega)  \\
-V^{2}G_{01} (\omega)& \omega-\epsilon-V^{2} G_{00}(\omega)
\end{array} \right)^{-1}\,,
\label{Gimp}
\end{equation}
where $G_{nm}(\omega)=\sum_{\bm{k}}G_{nm}(\omega,\bm{k})$ and   $G_{nm}(\omega,{\bm{k}})=\langle\langle a_{n\bm{k}}^{},a_{m\bm{k}}^{\dag }\rangle\rangle$. Explicit expressions for the latter propagators are
 \begin{equation} \label{G00}
 G_{00}(\omega,{\bm{k}})\smeq\frac{\omega(\omega\smmi \hbar\Omega)[\omega(\omega\smmi \hbar\Omega)^2-\omega|\phi_{\bm{k}}|^2-(\omega\smmi \hbar\Omega)|A_{\bm{k}}|^2]}{D(\omega,\bm{k})}\,,
\end{equation}
and
\begin{equation}
 G_{01}(\omega,{\bm{k}})=\frac{\omega(\omega- \hbar\Omega)[(\omega- \hbar\Omega)\phi_{\bm{k}}+\omega\phi_{\bm{k}}^*  ]A_{\bm{k}}^*}{D(\omega,\bm{k})}\,,
 \label{G01}
\end{equation}
with 
\begin{eqnarray}
D(\omega,\bm{k})&=&[(\omega^2-|\phi_{\bm{k}}|^2)(\omega- \hbar\Omega)-\omega|A_{\bm{k}}|^2]\\
\nonumber
&&[((\omega-\hbar\Omega)^2-|\phi_{\bm{k}}|^2)\omega-(\omega-\hbar\Omega)|A_{\bm{k}}|^2]\\
\nonumber
&&-[(\omega-\hbar\Omega)\phi_{\bm{k}}+\omega\phi_{\bm{k}}^*][\omega \phi_{\bm{k}}+(\omega-\hbar\Omega)\phi_{\bm{k}}^*]|A_{\bm{k}}|^2\,.
\end{eqnarray}
The propagator $G_{11}({\omega,\bm{k}})$ can be obtained from  $G_{00}(\omega,{\bm{k}})$ by the substitution $\omega \leftrightarrow(\omega-\hbar\Omega)$ while $G_{10}^{r}(\omega,{\bm{k}})=G_{01}^{a}(\omega,{\bm{k}})^*$ where $r$ and $a$ denote retarded and advanced, respectively.

The energies of the bound states (if they exist) are determined by the poles of the trace of Eq. (\ref{Gimp}).
This can be found numerically (as it is done below) but to grasp the main physical ingredients it is better to analyze the problem perturbatively.
The imaginary part of the retarded self-energy $V^{2} G_{00}^r(\omega)$ is proportional to the LDOS of the irradiated pristine graphene projected onto the $m=0$ Floquet subspace and has a dynamical gap centered at $\hbar\Omega /2$. Its real part, on the other hand, is non zero inside the gap and diverges at the gap edges with different signs on each edge. 
As a consequence, to the lowest order in the impurity hybridization,  the impurity spectral density ($\propto-\mathrm{Im}(\mathcal{G}^r_{00}(\omega))$) has always a pole within the dynamical gap with an energy given by $\omega-\epsilon-V^{2} G_{00}^r(\omega)=0$.  
Assuming, for the sake of argument, that $\epsilon=0$ , it is easy to see that in the same order and in the $m=1$ Floquet subspace there is a bound state symmetrically positioned with respect to the gap center. 
\begin{figure}[t]
\includegraphics[width=1.\columnwidth]{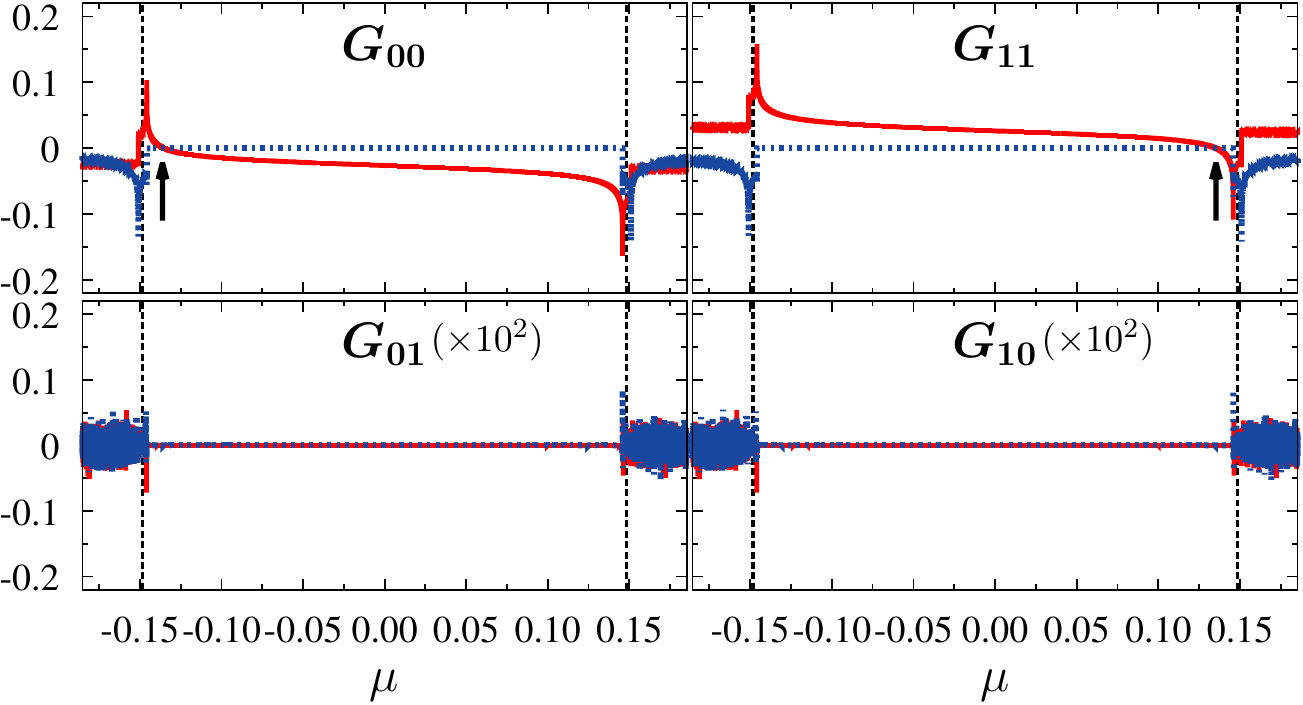}
\caption{(color online) Local retarded Green's functions ($G_{nm}$) for the irradiated pristine graphene corresponding to the Floquet subspaces $m,n=0,1$. These calculations were obtained numerically by decimation procedures, projecting onto only one C atom and using the same parameters $\eta$ and $\hbar\Omega$ as earlier. The black arrows show the zeroes of $G_{00}$ and $G_{11}$, i.e., the quasi-energies of the bound states for the vacancy (see main text).}  \label{Greens} 
\end{figure}

These results are in fact exact since $G_{01}(\omega)=G_{10}(\omega)=0$ within the dynamical gap---we checked this numerically (see Fig. \ref{Greens}) but it can also be obtained from Eq. (\ref{G01}) in the low energy limit where $\phi_{\bm{k}}$ ($A_{\bm{k}}$, $D(\omega,\bm{k})$) is odd (even) under the change $\bm{k}\rightarrow-\bm{k}$. Therefore, there are two bound states, belonging to the  $m=0$ and $m=1$ Floquet replicas, whose energies are given by the zeroes of $\omega - \epsilon - V^2 G_{00}(\omega)$ and $\omega - \epsilon - \hbar \Omega - V^2 G_{11}(\omega)$, respectively. 

\begin{figure}[t]
\includegraphics[width=0.99\columnwidth]{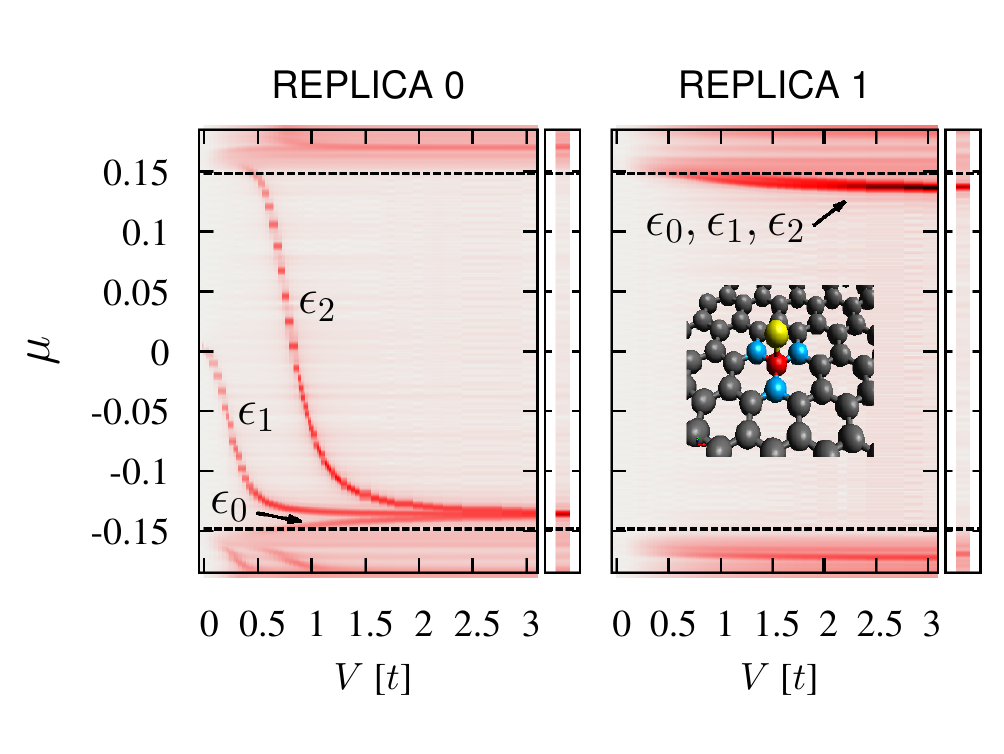}
\caption{(color online) Floquet local density of states around the center of the dynamical gap as a function of hybridization strength, projected on the three first neighbors of the carbon atom at which the impurity is adsorbed. Left and right panels correspond to projections onto $m=0$ and $m=1$ Floquet subspaces, respectively. The bound states inside the dynamical gap were obtained separately for adatoms with single orbital energies: $\epsilon_0=0$, $\epsilon_1= \hbar \Omega/2$ and $\epsilon_2= \hbar \Omega$.  At the right of each panel we show the vacancy limit where both the adatom and the C atom below it  are removed (red an yellow atoms in the inset).}  \label{adatom} 
\end{figure}

Figure \ref{adatom} shows a color map of the local Floquet spectral density (corresponding to the three sites around the adatom) calculated using the Chebyshev method, described in Sec. \ref{TBM},  as a function of the hybridization matrix element $V$ for different values of $\epsilon$. 
We found that while the  energies of the bound states depend on the energy of the adatom, these states are always present regardless of the size of the hybridization. The symmetry between replicas is broken if $\epsilon\neq0$ and it is only recovered in the limit of very large hybridization where the problem reduces to that of a vacancy. In this vacancy limit ($V\rightarrow\infty$), the position of the bound states, are given by the solution of $G_{00}^r(\omega)=0$ and $G_{11}^r(\omega)=0$ (indicated by the arrows in Fig. \ref{Greens}),
being the spectrum within the dynamical gap symmetric with respect to the gap center.

Interestingly, when looking at the weight of each of these states on the adatom and the three carbon atoms around it, one finds that they belong to a single replica. This particular result is a consequence that the coupling between the adatom and the layer of graphene was considered unaffected by the radiation field--- see Figure \ref{adatom}.\\
 
\section{Conclusions}
\label{conclu}
In summary, we have presented a detailed study of the Floquet bound states associated to defects in graphene illuminated by a laser. In particular, we focus on the bound states at the dynamical gap ($\hbar\Omega/2$) using both analytical and numerical techniques applied to different defect types. 

On one hand we consider large hole-like defects with different terminations. In this case, we show how the number of bound states increases with the defect radius and that the spectrum depends on the shape and type of lattice termination. In the case of cZZBC we proved analytically that in the limit of large radii the discrete bound states can be seen as nanoribbon-like chiral states\cite{Perez-Piskunow2014,Usaj2014a} with a quantized linear  quasi-momentum, as might have been anticipated.  Staggered like potential (infinity mass boundary conditions) was also discussed with similar results, except that in this case there is a clear distinction between the two Dirac cones, and only one of them support chiral bound states.
The chiral nature of the states was corroborated by an explicit calculation of the probability currents around the defect in the two analytical cases we presented.

On the other hand, we also consider point-like defects such as vacancies and adatoms and show that they also exhibit bound states around them. While the bound states spectrum depends on the value of the adatoms' orbital energy ($\epsilon$) in the large hybridization or vacancy limit, they remain close to the bottom (top) border of the gap in the $m=0$ ($m=1$) replica. 

Following the argument presented in Ref. [\onlinecite{Perez-Piskunow2015}] one can anticipate that additional  bound states will also appear inside the high order gaps induced by high order photon processes. The contribution of such states to the spectral density projected onto the $m=0$ replica is parametrically smaller provided $\eta\ll 1$. 

It remains a challenge for future work to evaluate the effect of these bound states on the bulk transport properties of dirty samples.

\section{Acknowledgements}
We acknowledge financial support from PICTs 2008-2236, 2011-1552 and
Bicentenario 2010-1060 from ANPCyT, PIP 11220080101821 and 11220110100832 from CONICET and 06/C415 SeCyT-UNC. GU and LEFFT acknowledge support from the ICTP associateship program, GU also thanks the Simons Foundation. LEFFT is on leave from CONICET and Universidad Nacional de C\'ordoba (Argentina).

\appendix

\section{Boundary conditions}
\label{BoundaryConditions_App}
As we already mentioned in Sec. \ref{BoundaryConditions}, an arbitrary BC can be imposed by knowing the matrix $\bm{M}$ and their action on the wavefunction evaluated at the boundary: $\Psi=\bm{M}\Psi$ [\onlinecite{McCann2004}]. 
It can be demonstrated that boundary conditions are determined by two unit vectors: $\hat{\bm{\nu}}$ acting on the isospin (valleys) and $\hat{\bm{n}}$ acting on the pseudospin (sublattices) [\onlinecite{Akhmerov2007}].
In the isotropic representation $\bm{M}= (\hat{\bm{\nu}} \cdot \bm{\tau})\otimes(\hat{\bm{n}} \cdot \bm{\sigma})$, where $\bm{\tau}$ and $\bm{\sigma}$ are the Pauli's matrices belonging to the isospin and pseudospin subspaces, respectively.
In the following, we show the explicit form of the matrix $\bm{M}$ for regular polygons, included the circle as the limit case, and the three kinds of BCs considered in this work.\\

\begin{figure}[t]
\centering
\includegraphics[width=0.99\columnwidth]{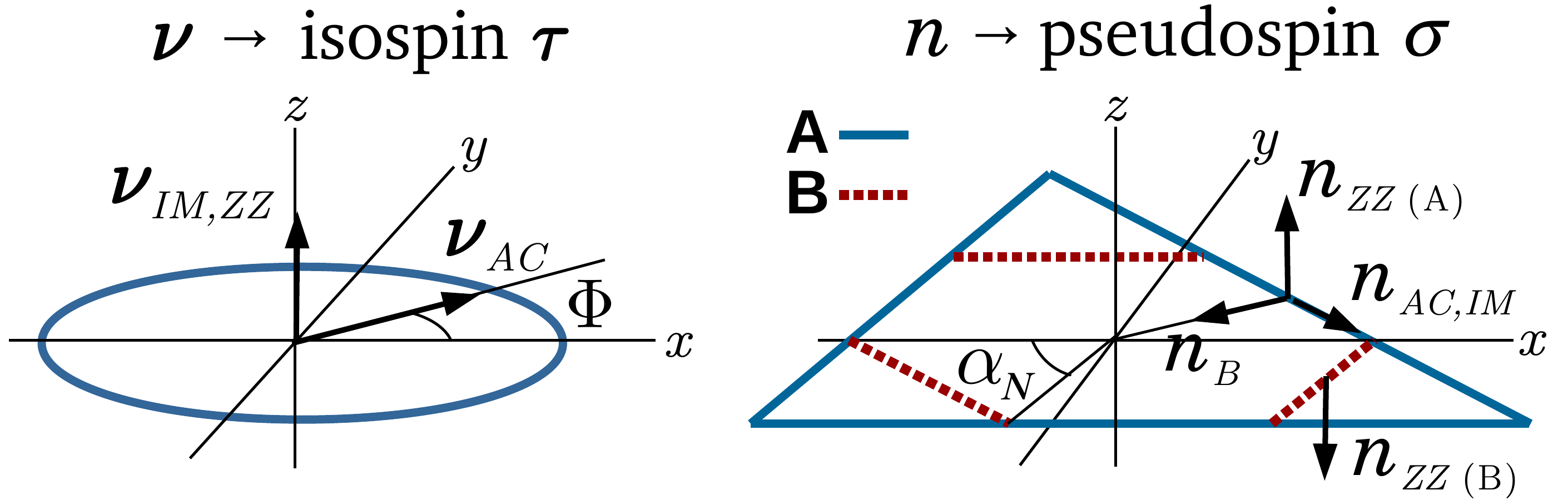}
\caption{(color online) Unit vectors $\hat{\bm{\nu}}$ and $\hat{\bm{n}}$ determine the boundary conditions. Each one acts on distinct degrees of freedom: $\hat{\bm{\nu}}$ acts on the isospin (valleys) and $\hat{\bm{n}}$ acts on the pseudospin (sublattices). Unit vector $\hat{\bm{n}}$ depends on the number of sides $N$ of the regular polygon via the normal unit vector $\hat{\bm{n}}_B$ for ABC and IMBC and the alternating nature of the sublattices terminations for ZZBC. This dependence also implies a dependence with the polar angle $\varphi$.  However, for a triangular defect there is only one type of sublattice termination (all atoms belong to the same sublattice, A sublattice in the right scheme) and the matrix $M$ in the Eq. (\ref{M_Ap})  becomes $(\varphi)-$independent.}
\label{iso-pseudo-spin}
\end{figure}

For both, ZZBC and ABC/IMBC, $\hat{\bm{n}}=\pm\hat{\bm{z}}$ (the sign depends on the sublattice termination) and  $\hat{\bm{n}}(\varphi)=\hat{\bm{z}}\times\hat{\bm{n}}_B(\varphi)$, respectively (see Fig. \ref{iso-pseudo-spin}).  
In the latter expression, $\hat{\bm{n}}_B(\varphi)$ is the normal unit vector located at the edges of the defects pointing outward from the region of interest---for our purpose, this unit vector pointing to the center of the defects.
For simplicity, we introduce the angle $\gamma_p$ related to the pseudospin degree of freedom. Thus,
we can handle both types of boundary conditions at the same time by writing
\begin{equation} \label{versor-n}
	\hat{\bm{n}}(\varphi)= \sin \gamma_p\,\hat{\bm{z}}\times\hat{\bm{n}}_B(\varphi) + \cos \gamma_p\, \hat{\bm{z}},
\end{equation}
and chose $\gamma_p=0(\pi)$ or $\gamma_p=\pi/2$ in order to select one or another type of BC.
It must be noted that while $z$-component is exclusively related with the ZZBC, the $xy$-components are related with ABC and IMBC--- the difference between two latter types of BCs resides in the isospin $\bm{\nu}$ i.e., in the details of the lattice terminations.
For a regular polygon with $N$ sides, the normal unit vector pointing inwards has the form
\bea \label{versor-nb}
	\hat{\bm{n}}_B (\varphi) &=& \sum_{j=1}^N \tilde{\Theta}_{j,N}(\varphi) \times \\
	&&  \left\{\cos\left[\alpha_N\left(j-1/2\right) \right] \hat{\bm{x}} + \sin\left[\alpha_N\left(j-1/2\right)\right] \hat{\bm{y}}\right\}, \nonumber
\eea
where $\tilde{\Theta}_{j,N}(\varphi)=\Theta(\varphi - j\,\alpha_N) - \Theta\left(\varphi - (j-1)\alpha_N\right)$, $\alpha_N=2\pi/N$, and $\Theta(\varphi)$ is the usual step function.
Using Eqs. (\ref{versor-n}) and (\ref{versor-nb}), we can write
\bea
	\hat{\bm{n}}\cdot\bm{\sigma} &=& 
\begin{pmatrix}
\cos \gamma_p & \Xi_N(\varphi)\,\sin\gamma_p \\
\Xi_N^*(\varphi) \,\sin\gamma_p & - \cos \gamma_p
\end{pmatrix}.
\eea
where $\Xi_N(\varphi) =  \sum_{j=1}^N i \tilde{\Theta}_{j,N}(\varphi)\, e^{-i \alpha_N (j-1/2)}$. It is useful to rewrite this quantity as a Fourier series
\be
 	\Xi_N(\varphi) = \sum_{m=-\infty}^{\infty} i\,A_{m,N} e^{i(mN-1)\varphi}
 		\label{xi}
\ee
where $A_{m,N} = \text{sinc}(\pi/N)/(1-mN)$ and  $\text{sinc}(x)=\sin{x}/x$. It is straightforward to see that for circular defects we have  $\lim_{N\to\infty} A_{m,N}=\delta_{m0}$.

Analogously, for the isospin degree of freedom: $\hat{\bm{\nu}}=\hat{\bm{z}}$ and $\hat{\bm{\nu}} \cdot \hat{\bm{z}}=0$ for the ZZBC/IMBC and ABC, respectively. Introducing now the angle $\gamma_i$, we can write this all three BCs in the form
\bea
	\hat{\bm{\nu}}\cdot\bm{\tau} &=& 
\begin{pmatrix}
\cos \gamma_i & e^{-i \Phi}\,\sin\gamma_i \\
e^{i \Phi}\,\sin\gamma_i & - \cos \gamma_i
\end{pmatrix}, \label{decoupled}
\eea
where $\gamma_i=0$ for both, ZZBC and IMBC---for these BCs $K$ and $K'$ cones are decoupled. For the ABC however, $\hat{\bm{\nu}}$ lies on the $xy$ plane, i.e., $\gamma_i=\pi/2$---the $\Phi$-phase is only relevant for the ABC, however, the analytic solutions of the ABC is out of the scope of this work.

Finally, the matrix $\bm{M}$ in terms of the angles $(\gamma_i,\gamma_p)$ is
\bea \label{M_Ap}
	&\bm{M}&(\varphi) = (\hat{\bm{\nu}}\cdot\bm{\tau})\otimes (\hat{\bm{n}}\cdot\bm{\sigma}) \\
 &=& \begin{pmatrix}
\cos \gamma_i & e^{-i \Phi}\,\sin\gamma_i \\
e^{i \Phi}\,\sin\gamma_i & - \cos \gamma_i
\end{pmatrix} \otimes \begin{pmatrix}
\cos \gamma_p & \Xi_N(\varphi)\,\sin\gamma_p \\
\Xi_N^*(\varphi) \,\sin\gamma_p & - \cos \gamma_p
\end{pmatrix},\nonumber
\eea
and the analogous to the set of conditions (\ref{BCeqs}), is 
\bea
\text{ZZBC}&\to& \gamma_i=0\,, \gamma_p = 0(\pi),\label{ZZBC_Ap} \nonumber \\
\text{ABC}&\to& \gamma_i=\pi/2\,, \gamma_p = \pi/2, \label{BCeqs_Ap}\\
\text{IMBC}&\to&\gamma_i=0\,,\gamma_p = \pi/2\,.\nonumber
\eea 
 
The dependence of $\bm{M}$ with the polar angle $\varphi$ relies on the pseudospin contribution.
Triangles and hexagons are the unique regular polygons with well defined zigzag terminations. Therefore, the angle $\gamma_p$ for the ZZBC can behave in two different ways: it can be constant along the boundary of the defect (triangular defects), or it can alternate between 0 and $\pi$ depending on the sublattice termination (hexagonal defects) (see Fig. \ref{iso-pseudo-spin}). 
In order to tackle circular defects with ZZBCs, one is tempted to define the circle case as the limit of a polygon with a  $N$ large enough and an alternating $\bm{n} = \pm \bm{z}$ on their faces, corresponding to different sublattices terminations. However, this artificial limit is misleading because of is not possible construct such a defect, i.e., a regular polygon with $N>6$ whose edges were constructed exclusively of zigzag neither armchair terminations.
For simplicity, throughout this article we only work with ZZBC for triangular defects, in such a way that $\bm{M}$ is $\varphi$-independent.
 In this case,  introduce the first condition of the set (\ref{ZZBC_Ap}) into Eq. (\ref{M_Ap}) leads to $\psi_{B,l}(\varphi,\xi_0)=\psi_{B,l}'(\varphi,\xi_0)=0$
---we emphasize that, in the isotropic representation, $\psi=(\psi_A,\psi_B,-\psi_B',\psi_A')^\text{T}$ must be used. Thus,
there are two equations per cone [Eqs. (\ref{cond_zzbc}) for the $K$ cone], one for each Floquet replica, which allow us to find the relation between coefficients $c_{+}$ and $c_{-}$, and then, the quasi-energies $\mu_l$ [solutions of the Eqs . (\ref{condKcone}) and (\ref{condKpcone})].

On the other hand, for the ABC and the IMBC, the dependence of matrix $\bm{M}$ with the polar angle $\varphi$ can not be avoided whatever the number of sides of the polygon considered. 
Even in the limit of circular defects: $\lim_{N \to \infty} \Xi_N (\varphi) = i e^{-i\varphi}$---unlike the ZZBC and the ABC, the circular defect is well defined for the IMBC because of this kind of BC does not depend on the details of the terminations at the edges (zigzag, armchair or mixing of them). As a consequence, for the IMBC the strategy to find the quasi-energies is quite different from that of the ZZBC (see App. \ref{imbc_sol_ap}).

\section{Solutions for the cZZBC and the IMBC - Circular defects.}
For the $K$ cone, the Floquet state restricted to $n=0$ and $n=1$ Floquet subspaces, has the form
\begin{equation}\label{Floquet_Ap}
	\Psi_l (\bm{r},t) = \frac{e^{-i \varepsilon_l t}}{\sqrt{N_l}} \begin{pmatrix}
	u_{1A,l}(\bm{r})\, e^{i\Omega t} + u_{0A,l}(\bm{r}) \\
	u_{1B,l}(\bm{r})\, e^{i\Omega t} + u_{0B,l}(\bm{r})
\end{pmatrix},
\end{equation}
where the components are
\begin{eqnarray} \label{componentes_Ap}
	\phi_{l}(\varphi,\xi) = \begin{pmatrix}
	u_{1A,l} \\
	u_{1B,l} \\
	u_{0A,l} \\
	u_{0B,l}
\end{pmatrix} = \begin{pmatrix} \frac{i}{1-\mu_l} e^{i(l-1)\varphi} \tilde{f}_l(\xi) \\
e^{il\varphi}	f_l(\xi) \\
 e^{il\varphi}g_l(\xi)  \\
	\frac{-i}{1+\mu_l} e^{i(l+1)\varphi} \tilde{g}_l(\xi)
\end{pmatrix} ,
\end{eqnarray}
Hence,
\begin{eqnarray} \label{definitions_Ap}
	f_l(\xi) &=& c_{+} K_l(\lambda_{+}\xi) + c_{-} K_l(\lambda_{-}\xi), \nonumber \\
 \tilde{f}_l(\xi) &=& -c_{+} \lambda_{+} K_{l-1}(\lambda_{+}\xi) - c_{-} \lambda_{-} K_{l-1}(\lambda_{-}\xi), \nonumber \\
	g_l(\xi) &=& d_{+} K_l(\lambda_{+}\xi) + d_{-} K_l(\lambda_{-}\xi), \nonumber \\
 \tilde{g}_l(\xi) &=& -d_{+} \lambda_{+} K_{l+1}(\lambda_{+}\xi) - d_{-} \lambda_{-} K_{l+1}(\lambda_{-}\xi),
\end{eqnarray}
where $\xi = k_0 r$, $\lambda_{\pm} = \sqrt{-1-\mu_l^2 \pm 2 \sqrt{-\eta^2 + \mu_l^2(1+\eta^2)}}$, $\beta_{\pm} = -\left[\lambda_\pm^2 + (1-\mu_l)^2\right]/\left[2\eta (1-\mu_l)\right]$ and $d_{\pm} = \beta_{\pm} c_{\pm}$. We also notice that $\lambda_{-}= \lambda_{+}^*$ and $K_{\nu}(z^*)=K_{\nu}^*(z)$.

Because of cZZBC and IMBC do not mix different valleys [see Eq. (\ref{decoupled})], we can impose normalization conditions for each valley in an independent way.
According to Eqs. (\ref{Floquet_Ap}), and the angular dependence of the components of the Floquet state given by (\ref{componentes_Ap}), the normalization constant results time-independent 
\begin{equation} \label{norma_Ap}
	N_l = \frac{2\pi}{k_0^2} \int_{\xi_0}^\infty \left(|f_l(\xi)|^2 + |g_l(\xi)|^2 + \frac{|\tilde{f}_l(\xi)|^2}{(1-\mu_l)^2} + \frac{|\tilde{g}_l(\xi)|^2}{(1+\mu_l)^2} \right)\,\xi\,d\xi.
\end{equation}
Defining following quantities 
\bea
	P_{\nu} &=& \int_{\xi_0}^\infty  K_\nu (z \xi)K_\nu (z^* \xi)\, \xi\, d\xi \nonumber \\
              &=& \frac{\xi_0}{2} \frac{{\rm Im}\{z K_{\nu-1}(z \xi_0) K_{\nu}(z^* \xi_0)\}}{{\rm Re}\{z\}{\rm Im}\{z\}}, \nonumber \\
Q_{\nu} &=& \int_{\xi_0}^\infty  K_\nu (z \xi)K_\nu (z \xi)\, \xi\, d\xi  \nonumber \\
 &=& \frac{\xi_0}{2z}\times \\
 & &\left(2\nu K_{\nu-1}(z \xi_0) K_{\nu}(z \xi_0) - z \xi_0 [K_{\nu}^2(z \xi_0) - K_{\nu-1}^2(z \xi_0)] \right),\nonumber
\eea
(where $\lim_{\xi\to\infty}K(z\xi)=0$, was used), we can write the normalization constant as follow
\bea
	N_l &=& \frac{2\pi}{k_0^2} \left( (1 + |\beta_{+}|^2) P_{l} + |\lambda_{+}|^2 P_{l-1} + |\beta_{+}|^2 |\lambda_{+}|^2 P_{l+1} \right. \nonumber \\
 &+& \left. {\rm Re}\left\{ e^{i\varphi} [(1 + \beta_{+}^2) Q_{l} + \lambda_{+}^2 Q_{l-1} + \beta_{+}^2 \lambda_{+}^2 Q_{l+1} \right\} \right).
\eea

Here, different boundary conditions only modify relations between coefficients:  $e^{i  \theta} = c_{-}/c_{+}$. While for the cZZBC $e^{i \theta} = -K_l(\lambda_{+} \xi_0)/K_l(\lambda_{-} \xi_0)$, for the IMBC $e^{i \theta} = -\omega_{+} \beta_{+}/(\omega_{-} \beta_{-})$, with $\omega_{\pm} = (1+\mu_l)K_l(\lambda_{\pm} \xi_0)+\lambda_{\pm}K_{l+1}(\lambda_{\pm} \xi_0)$.\\

In order to obtain solutions belonging to the $K'$ cone, the isotropic representation requires that: $\psi_{A,l} \to -\psi_{B,l}'$ and $\psi_{B,l} \to \psi_{A,l}'$. By doing these replacements, the same procedure applied in Sec. \ref{model} leads to a set of equations analogous to Eqs. (\ref{eqreduced})---and their respective boundary condition $\psi_B'(\xi_0)=0$--- which, in principle, must be solved again. However, for the cZZBC case, latter set of equations and their boundary conditions can be obtained from that of belonging to the $K$ cone by doing follow changes: $(\mu,l)\to (-\mu,-l)$. Doing so, for the $K'$ cone we have
\begin{eqnarray} \label{componentesu2}
	\phi_{-l}' &=& \begin{pmatrix}
	u'_{1A,-l} \\
	u'_{1B,-l} \\
	u'_{0A,-l} \\
	u'_{0B,-l}
\end{pmatrix} = \begin{pmatrix}
	-e^{-il\varphi}	g_{l}(\xi) \\
	\frac{i}{1+\mu_{l}} e^{-i(l+1)\varphi} \tilde{g}_{l}(\xi) \\
   \frac{-i}{1-\mu_{l}} e^{-i(l-1)\varphi} \tilde{f}_{l}(\xi)   \\
 -e^{-il\varphi}f_{l}(\xi)
\end{pmatrix}.
\end{eqnarray}
The time averaged probability density current (over one period) only has an angular component as it is shown in Sec. \ref{chiral_current}. Then, the density currents for both cones are
\bea
	J_{l} &=& -2\,\text{Im}\{e^{i\varphi}\left(u_{1A,l}u_{1B,l}^*+u_{0A,l}u_{0B,l}^*\right)\}, \label{currentK}\\
	J_{l}' &=& -2\,\text{Im}\{e^{-i\varphi}\left(u_{1A,l}' u_{1B,l}'^* + u_{0A,l}' u_{0B,l}'^*\right)\}. \label{currentKp}
\eea
Hence, it is straightforward to see that $J_{l}=J_{-l}'$.

On the other hand, there is no any transformation between the $K$ and $K'$ cones for the IMBC case which simultaneously leaves invariant the set of differential equations and their respective boundary condition. Therefore, the set of quasi-energies for the $K'$ cone must be founded following the same procedure used for the $K$ cone. The isotropic representation imposes that
\begin{eqnarray} \label{componentesu3}
	\phi_{l}' &=& \begin{pmatrix}
	u'_{1A,l} \\
	u'_{1B,l} \\
	u'_{0A,l} \\
	u'_{0B,l}
\end{pmatrix} = \begin{pmatrix}
	e^{il\varphi}	f_{l}(\xi) \\
   \frac{-i}{1-\mu_{l}} e^{i(l-1)\varphi} \tilde{f}_{l}(\xi)  \\
	\frac{-i}{1+\mu_{l}} e^{i(l+1)\varphi} \tilde{g}_{l}(\xi) \\
 -e^{il\varphi}g_{l}(\xi)
\end{pmatrix}.
\end{eqnarray}
For the $K'$ cone, coefficients $c_{+}$ and $c_{-}$ are related now by the phase $e^{i \theta} = \omega'_{+} \beta_{+}/(\omega'_{-} \beta_{-})$, with $\omega'_{\pm} = (1+\mu_l)K_l(\lambda_{\pm} \xi_0)-\lambda_{\pm}K_{l+1}(\lambda_{\pm} \xi_0)$. 
The time averaged probability density currents for each cone are also given by Eqs. (\ref{currentK}) and (\ref{currentKp}). Nevertheless, there is no any relation between $J_{l}$ and $J_{l}'$. \\ 

\begin{figure}[t!]
\centering
\includegraphics[width=0.95\columnwidth]{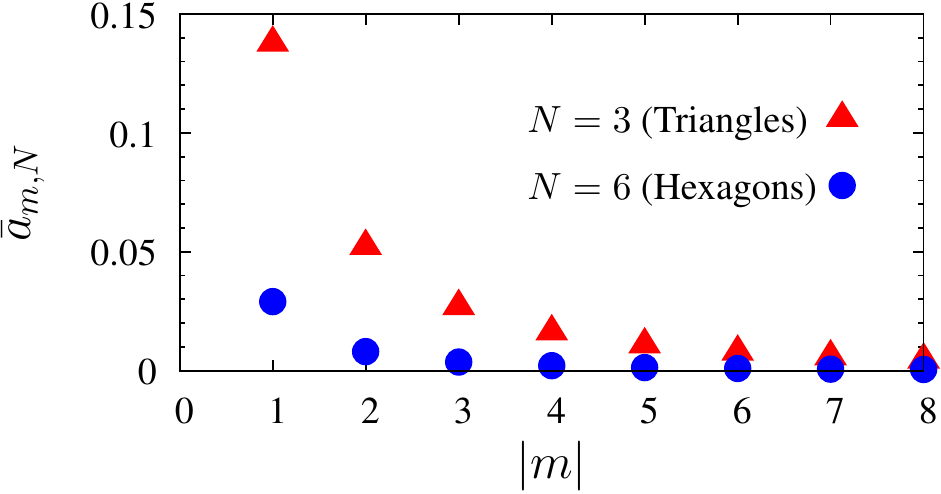}
\caption{(color online) 
Relative contributions, measured by $\bar{a}_{mN,N}=a_{mN,N}/a_{0,N}$, of higher orders in the $N\varphi$-dependence to the zero-order expansion in the Eq. (\ref{desarrollo-R}) for triangular and hexagonal defects.}
\label{a_coeff}
\end{figure}

\section{Solutions for the IMBC - Polygonal defects.} \label{imbc_sol_ap}

For simplicity, we will only tackle the IMBC, which does not mix cones. In this case,  introducing the third condition of the set (\ref{BCeqs_Ap}) into Eq. (\ref{M_Ap}) leads to a mixing of solutions with different $l$ quantum numbers due to the aforementioned dependence, i.e.
\bea
\nonumber
\sum_l f_{l}(\xi_0) e^{il\varphi} \mp \sum_{l'} \frac{i \tilde{f}_{l'}(\xi_0)}{1-\mu} \Xi_N^*(\varphi)  e^{i(l'-1)\varphi} &=& 0\,,\\
\sum_l g_{l}(\xi_0) e^{il\varphi} \pm \sum_{l'} \frac{i \tilde{g}_{l'}(\xi_0)}{1+\mu} \Xi_N(\varphi)  e^{i(l'+1)\varphi} &=& 0,
\label{imbc_syseq}
\eea
where the upper (lower) sign refers to the $K(K')$ cone and components given by Eq. (\ref{componentes_Ap}), [(\ref{componentesu2})] were used.
We also have to account for the dependence of the coordinates of the edges with the polar angle $\varphi$, i.e. $\xi_0(\varphi)$.
For regular polygons with $N$ sides, the points located at the edges can be written as 
\be
 	R(\varphi) = R_0 \sum_{m=-\infty}^{\infty} a_{m,N} e^{i mN\varphi},
 		\label{desarrollo-R}
\ee
where coefficients $a_{m,N}$ are given by
\be
 	a_{m,N} =  N \int_{-\alpha_N/2}^{\alpha_N/2} \frac{e^{-i m N \phi}}{\cos{\phi}} d\phi,
 		\label{coeff-R}
\ee
$R_0$ is the apothem of the polygon and $\bar{R}_0 = R_0\, a_{0,N}$ represents the mean value of their radii. For triangles and hexagons, $a_{0,3} = 3\ln(2+\sqrt{3})/\pi$ and $a_{0,6} = 3\ln 3/\pi$, respectively.
In the large $N$ limit, the deviations of $R(\varphi)$ with respect to $\bar{R}_0$ are small and we can expand the modified Bessel functions of second kind $K_{\nu}$ appearing in Eqs. (\ref{imbc_syseq}) to first order on the deviation. That is, 
\be
K_{\nu}\left(\lambda\xi_0(\varphi)\right) \simeq K_{\nu}(\lambda\bar{\xi}_0) + \frac{\partial K_{\nu}(\lambda \xi_0)}{\partial \xi_0}|_{\bar{\xi}_0} \left(\xi_0(\varphi)-\bar{\xi}_0\right) \,.
\ee 
Using this approximation and Eq. (\ref{xi}), the conditions given by Eqs. (\ref{imbc_syseq}) can be rewritten as
\bea
\nonumber
 &f_{l}&(\bar{\xi}_0) + \sum_{m\neq0}  \bar{a}_{m,N}f'_{l-mN}(\bar{\xi}_0) \bar{\xi}_0 = \\ 
 &=& \pm \sum_{s} \frac{A_{s,N}}{1-\mu}\left(\tilde{f}_{l+sN}(\xi_0) + \sum_{n\neq0} \bar{a}_{n,N}\tilde{f}'_{l-nN+sN}(\bar{\xi}_0) \bar{\xi}_0 \right),\nonumber \\
 &g_{l}&(\bar{\xi}_0) + \sum_{m\neq0}  \bar{a}_{m,N}g'_{l-mN}(\bar{\xi}_0) \bar{\xi}_0 = \label{imbc_syseq_2} \\ 
 &=& \pm \sum_{s} \frac{A_{s,N}}{1+\mu}\left(\tilde{g}_{l-sN}(\xi_0) + \sum_{n\neq0} \bar{a}_{n,N}\tilde{g}'_{l-nN-sN}(\bar{\xi}_0) \bar{\xi}_0 \right),\nonumber
\eea
where  $f'$ ($g'$) indicates the first derivative with respect to $\xi$ of $f$ ($g$) and $\bar{a}_{m,N} = a_{m,N}/a_{0,N}$. Coefficients $\bar{a}_{m,N}$ are even functions of $m$ and they vanish quickly as $m$ grows (see figure).

It is straightforward to see that only for circular defects, the mixing among different $l$ quantum numbers is removed, since $\lim_{N\to\infty} a_{m\neq0,N} = \lim_{N\to\infty}A_{m\neq0,N} =0$ and $\lim_{N\to\infty} a_{0,N}=\lim_{N\to\infty} A_{0,N}=1$.

Finally, in order to find the quasi-energies, the infinite series in the Eqs. (\ref{imbc_syseq_2}) must be truncated. Doing so, it is possible to write a system with $2d$ equations for $d$ quasi-energies (each quasi-energy introduce two additional coefficients: $c_{+}$ and $c_{-}$) and then, find their solutions.

\begin{figure}[t]
\includegraphics[width=0.9\columnwidth]{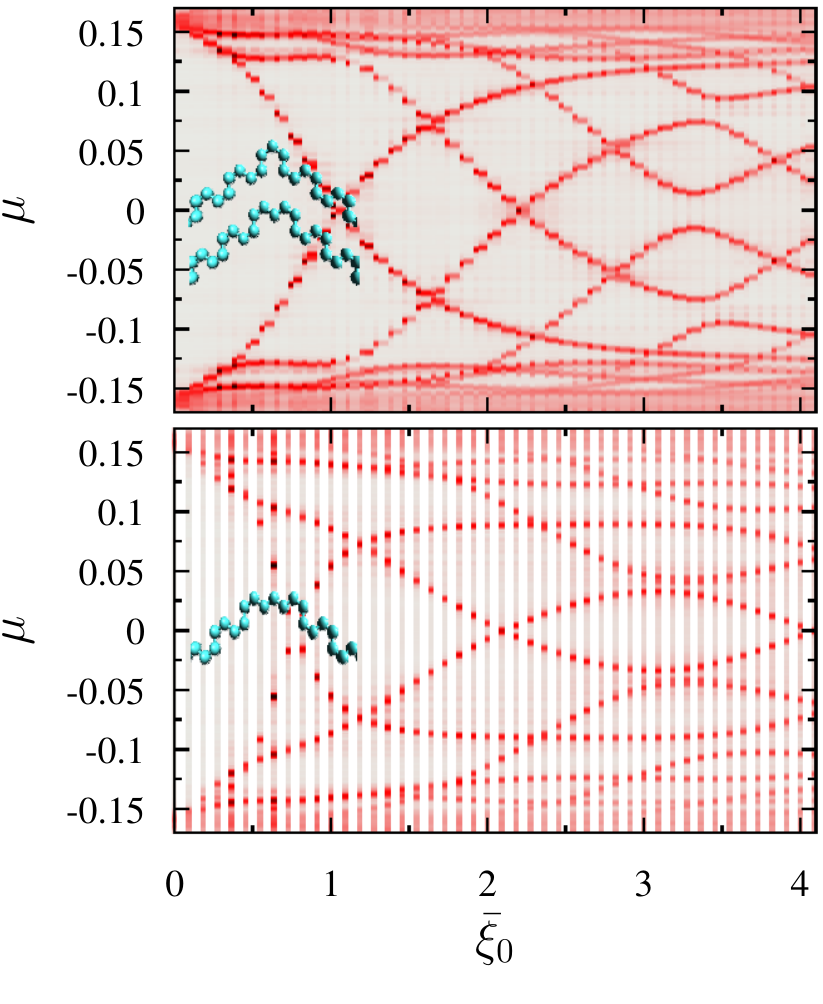}
\caption{(color online)  Same as Fig. \ref{TH-hole} for the three possible distinct hexagonal hole defects. The FLDOS in the top panel is the same showed in the Fig. \ref{TH-hole} for this kind of defects. The FLDOS for remaining configuration is showed in the bottom panel. It depends on the microscopic details beyond the zigzag or armchair terminations.}  
\label{H-hole_append} 
\end{figure}

\section{FLDOSs for hexagonal configurations.}
 \label{hex_conf}
Hexagonal defects with armchair terminations show only three possible distinct configurations. Even when all these three configurations have the same armchair terminations along their edges, they differ in the way  their sides match at the vertices. As already was mentioned in Sec. \ref{TBM}, the FLDOS for staggered potential defects are independent of the microscopic details as end terminations. However, the FLDOS for hexagonal hole defects does depend on the latter ones showing two different behaviors. We are only  interested in those configurations whose FLDOSs can be understood in terms of the wave functions for the low energy model and the boundaries conditions studied in Sec.   \ref{model} and  Sec. \ref{BoundaryConditions}, respectively.
In the top panel of the Fig. \ref{H-hole_append}  we show the FLDOS for two such configurations (see diagram left on it). The FLDOS for the remaining configuration is shown in the bottom panel. The FLDOS for the latter one is perturbed by microscopic details and it is beyond the scope of the present work.

%
\end{document}